\documentclass[aps,prx,,twocolumn,notitlepage,amsfonts,citeautoscript,superscriptaddress,showpacs,letter]{revtex4-2}

\begingroup\makeatletter
\catcode`,=\active
\global\let\breqn@comma,
\protected\gdef,{\ifmmode\expandafter\breqn@comma\else\expandafter\active@comma\fi}
\endgroup

\usepackage{graphicx,epstopdf,color}
\usepackage{amsmath,amssymb,mathrsfs}
\usepackage{hyperref}
\usepackage[dvipsnames]{xcolor}
\usepackage{url}
\usepackage{soul}
\usepackage{cancel}
\usepackage{cleveref}
\usepackage{bm}
\usepackage{xr}
\usepackage[caption=false]{subfig}
\usepackage{epstopdf}

\crefname{equation}{Eq.}{Eqs.}
\Crefname{equation}{Equation}{Equations}
\crefname{figure}{Fig.}{Figs.}
\Crefname{figure}{Figure}{Figures}
\crefname{section}{Sec.}{Secs.}
\Crefname{section}{Section}{Sections}
\crefname{appendix}{Appendix}{Apps.}
\Crefname{appendix}{Appendix}{Apps.}
\crefname{paragraph}{Sec.}{Secs.}
\crefname{table}{Table}{Tables}

\graphicspath{{nb_fig/}}
\newcommand{\hH}{\hat{H}}
\newcommand{\ha}{\hat{a}}

\renewcommand{\Re}{\text{Re}}

\makeatletter
\def\maketitle{
\@author@finish
\title@column\titleblock@produce
\suppressfloats[t]}
\makeatother

\begin{document}
\title{Non-perturbative switching rates in bistable open quantum systems:\\ from driven Kerr oscillators to dissipative cat qubits}
\author{L{\ifmmode\acute{e}\else\'{e}\fi}on Carde}
\thanks{leon.carde@alice-bob.com}

\affiliation{Laboratoire de Physique de l'\'{E}cole Normale Sup\'{e}rieure$,$ Inria$,$ ENS$,$ Mines ParisTech$,$ Universit\'{e} PSL$,$ Sorbonne Universit\'{e}, Paris$,$ France}
\affiliation{Alice and Bob$,$ 53 Bd du G\'{e}n\'{e}ral Martial Valin$,$ 75015$,$ Paris$,$ France}
\author{Ronan Gautier}
\affiliation{Alice and Bob$,$ 53 Bd du G\'{e}n\'{e}ral Martial Valin$,$ 75015$,$ Paris$,$ France}
\author{Nicolas Didier}
\affiliation{Alice and Bob$,$ 53 Bd du G\'{e}n\'{e}ral Martial Valin$,$ 75015$,$ Paris$,$ France}
\author{Alexandru Petrescu}
\affiliation{Laboratoire de Physique de l'\'{E}cole Normale Sup\'{e}rieure$,$ Inria$,$ ENS$,$ Mines ParisTech$,$ Universit\'{e} PSL$,$ Sorbonne Universit\'{e}, Paris$,$ France}
\author{Joachim Cohen}
\affiliation{Alice and Bob$,$ 53 Bd du G\'{e}n\'{e}ral Martial Valin$,$ 75015$,$ Paris$,$ France}
\author{Alexander McDonald}
\affiliation{Institut Quantique and D\'{e}partement de Physique\protect\\
Universit\'{e} de Sherbrooke, Sherbrooke~J1K~2R1~Qu\'{e}bec, Canada}

\date{\today}
\begin{abstract}
    In this work, we use path integral techniques to predict the switching rate in a single-mode bistable open quantum system. While analytical expressions are well-known to be accessible for systems subject to Gaussian noise obeying classical detailed balance, we extend this approach to a class of open quantum systems, those which satisfy the recently-introduced notion of hidden time-reversal symmetry~\cite{roberts_hidden_2021}. In particular, in the context of quantum computing, we obtain analytical estimates of bit-flip error rates in cat-qubit architectures. We confirm these findings by comparing our results to numerically exact diagonalization of the Lindbladian. Our results provide a path towards exploring switching phenomena in multistable open quantum systems. 
\end{abstract}
\maketitle
\textit{Introduction}.---Bosonic quantum systems whose mean field dynamics exhibit bistability are central to many areas of quantum science and technology. For example, the driven-dissipative Kerr-oscillator is ubiquitous in several physical platforms such as the paradigmatic transmon qubit~\cite{Koch_Transmon_2007} in circuit quantum electrodynamics, optomechanical systems \cite{Bose_PRL_1997, Ludwig_PRL_2012}, and atomic ensembles \cite{Gupta_PRL_2007}. Further, dissipative and Kerr cat qubits ~\cite{mirrahimi_dynamically_2014, leghtas_confining_2015, lescanne_exponential_2020_short, puri_engineering_2017, grimm_stabilization_2020_format} leverage bistability to encode information in a two-dimensional subspace spanned by large coherent states of an oscillator. In these bistable systems, fluctuations will cause the system to switch from one stable state to the other. The rate at which this occurs, known as the switching rate, is of crucial importance in such systems. It is, for example, of fundamental and practical interest in driven-dissipative Kerr oscillators \cite{Siddiqi_PRL_2004, Muppalla_PRB_2018, Andersen_PRApplied_2020, dykman2012fluctuating, sepulcre_analytical_2025, castillo_experimental_2025}. In the context of cat qubits, it corresponds to the bit-flip rate, and is exponentially suppressed with the number of photons. ~\cite{lescanne_exponential_2020_short, RegladeBocquet_Nature2024, MarquetPRX2024, gautier_combined_2022_format, putterman_hardware_2025_short}. This results in a strong bias of the qubit noise that can be leveraged for fault-tolerant quantum computing~\cite{guillaud_repet_2019, Puri_sciavd_2020, XZZX_Puri_Darmawan_2021, chamberland_building_2022, ruiz_ldpc_2024}, and underscores the importance of understanding the switching rate.

Calculating this rate analytically has proven to be difficult~\cite{marthaler_switching_2006,mirrahimi_dynamically_2014,thompson_qubit_2022,lee_arealtime_2024, thompson_population_2025, boness_zero_2025,su_unraveling_2024_format,dubovitskii_bit-flip_2025}. Various methods have been used to do so, such as perturbative expansions in the Lindblad superoperator ~\cite{mirrahimi_dynamically_2014,regent_rouchon_2023, dubovitskii_bit-flip_2025} and semi-classical approximations~\cite{Andersen_PRApplied_2020,Peano_NJP_2014, Dykman_SP_1988,Dykman_PRE_2007,Lin_PRE_2015, marthaler_switching_2006,boness_zero_2025, su_unraveling_2024_format}. Resorting to path integral techniques to non-perturbatively obtain the switching rate, it is controlled by the most probable escape path connecting the two stable states~\cite{kamenev_field_2011, lee_arealtime_2024}. In that context, closed-form formulas have been derived for bistable open quantum systems in specific cases only~\cite{thompson_qubit_2022, thompson_population_2025}.
In contrast, for classical bistable systems subject to weak Gaussian white noise and satisfying detailed balance, the properties are well understood: the vanishing of the probability current imposes stringent constraints on the steady state~\cite{gardiner_handbook_2009} and the most probable escape path from a metastable state is the time-reversed version of the noise-free evolution~\cite{maier_escape_1993,kamenev_field_2011}.

\begin{figure}[t!]
    \centering
    \includegraphics[width=0.8\linewidth]{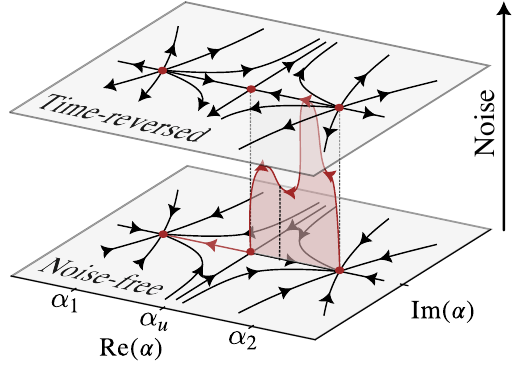}
    \caption{Schematic representation of a switching path in a dissipative bistable system.
     The lower plane represents the noise-free evolution and corresponds to the mean-field motion of the system, see~\cref{eq:mean-field_eq}. Here we consider the cat-state stabilization scenario where there are three fixed points (red-dots), two of which are stable and one is unstable within that manifold. In the extended Keldysh phase-space this manifold corresponds to the noise-free $b_{\rm q} = 0$ plane, see \cref{eq:b_q_def}. There exists a second manifold in which the evolution is reversed, see~\cref{eq:TR_parametrization}. This manifold coincides with the noise-free plane at the fixed points. Using this manifold we can, in this extended space, construct a path (red) connecting two mean-field stable fixed points.}
    \label{fig:drawing}
\end{figure}

In this Letter, we show that the switching rate of a large class of bistable single-mode bosonic open quantum systems can be analytically determined. These systems satisfy hidden-time reversal symmetry (HTRS), a recently-introduced quantum analogue of classical detailed balance~\cite{roberts_hidden_2021}. Remarkably, we demonstrate that the switching rates in these quantum models are determined in a manner that parallels their classical counterparts: they are controlled by the time-reversed noise-free paths, see \cref{fig:drawing}. Crucially, in these quantum models, this is only valid in a new coordinate system as we discuss below. We show that, for the systems we consider, the HTRS ensures a close-formed expression for the switching rate that is determined by the steady-state potential function in the complex-P representation~\cite{BartoloPRA_2016,roberts_hidden_2021}. We validate our analytical result against numerical diagonalization of the Lindbladian. Our work opens the door to exploring switching phenomena and multistability in other non-equilibrium quantum platforms, ranging from coupled cavities~\cite{Roberts_PRL_2023_bosons}, optomechanical arrays, and dissipative spin ensembles~\cite{Roberts_PRL_2023,Yao_PRL_2025}.

\textit{Setup}.--- We consider a single bosonic mode with HTRS~\cite{roberts_driven_2020} described by the Lindblad master equation ($\hbar = 1$) 
\begin{equation}\label{eq:Lindblad_Eq}
    \partial_t \hat{\rho} = \mathbf{L}( \hat{\rho}) 
    = -i[\hat{H}, \hat\rho] + \kappa_1 \mathbf{D}_{\ha}(\hat\rho) + \kappa_2 \mathbf{D}_{\ha^2}(\hat\rho)
\end{equation} 
where the Hamiltonian takes the form
\begin{equation}
\begin{split}
    \hH &= \Delta \ha^\dagger \ha - \frac{K}{2} \ha^{\dagger2} \ha^2 \\
    &\quad + \left(\Lambda_1 \ha^\dagger + \Lambda_2\ha^{\dagger2}+ \Lambda_3 \ha^{\dagger2}\ha + \text{h.c.}\right)
\end{split}
\end{equation}
where $\mathbf{D}_{\hat{X}}\hat{\rho} = \hat{X} \hat{\rho} \hat{X}^\dagger -\{\hat{X}^\dagger \hat{X}, \hat{\rho}\}/2$ is the dissipation superoperator. Here, the bosonic creation and annihilation operators obey the canonical commutation relation $[\ha,\ha^\dagger]=1$, $\kappa_1$ and $\kappa_2$ are single- and two-photon decay rates, $\Delta$ is the detuning, $K$ is the Kerr non-linearity, and $\Lambda_1$, $\Lambda_2$ and $\Lambda_3$ are single, two and cubic photon drives, respectively.

Our starting point is the mean-field equation for the expectation value of the annihilation operator $ \langle\ha(t)\rangle = \alpha(t)$, which reads
\begin{equation}\label{eq:mean-field_eq}
\begin{split}
    i\partial_t \alpha &= \alpha K_1 - \alpha |\alpha|^2 K_2 \\
    &\quad +\Lambda_{1} + 2\alpha^*\Lambda_2 +\alpha^2 \Lambda_3^* + 2 |\alpha|^2 \Lambda_3
\end{split}
\end{equation}
where $K_1 = \Delta - i\frac{\kappa_1}{2}$ and $K_2 = K + i\kappa_2$. Depending on the parameter regime, \cref{eq:mean-field_eq} can have multiple fixed points $\partial_t \alpha = 0$~\cite{venkatraman_driven_2024,meaney_quantum_2014,SM}. We limit ourselves to the parameter regimes where there are three fixed points, two stable denoted by $\alpha_{1}, \alpha_2$, and one unstable $\alpha_u$. We further assume that all three fixed points are well-separated in phase space. This parameter regime, for instance, includes the paradigmatic dissipative Kerr oscillator~\cite{lee_arealtime_2024, Andersen_PRApplied_2020} and dissipative Kerr cats~\cite{dubovitskii_bit-flip_2025, mirrahimi_dynamically_2014, su_unraveling_2024_format, boness_resonant_2024, boness_zero_2025}. 

As shown schematically in \cref{fig:drawing}, the mean-field equations separate phase space into two basins of attraction for the two stable fixed points. However, fluctuations give rise to trajectories that can switch between the two stable fixed points. This finite switching rate is the slowest decay rate of the system, and controls the dissipative gap of the associated Lindbladian. For dissipative cat qubits, this switching rate corresponds to the bit-flip rate~\cite{gautier_combined_2022_format,leregent_adiabatic_2024}.

\textit{Keldysh action and saddle-point equations}.--- To calculate this rate we use Keldysh field theory~\cite{sieberer_keldysh_2016, kamenev_field_2011}. Given the Lindbladian $\mathbf{L}$ in \cref{eq:Lindblad_Eq}, an associated Keldysh action functional can be constructed $S[a_{\rm cl}, a_{\rm q}] = \int dt \left(a_{\rm q}^*i \partial_t a_{\rm cl} -  a_{\rm q}i\partial_t a_{\rm cl}^* - i\mathcal{L}[a_{\rm cl}, a_{\rm q}]\right)$ with $a_{\rm cl}$ the phase-space variable, $a_{\rm q}$ the noise variable, and $\mathcal{L}$ the Lindbladian density functional given in \cite{SM}. The probability of tunneling from $\alpha_i$ to $\alpha_j$ for distinct $i,j \in \{1,2\}$ can be written as $\int_{
\substack{
a_{\rm cl}(- t/2) =  \alpha_{i} 
}
}^{a_{\rm cl}(t/2) =  \alpha_{j} }
\mathcal D[a_{\rm cl}, a_{\rm q}]
e^{i S[a_{\rm cl}, a_{\rm q}]}$
with $\mathcal{D}[a_{\rm cl}, a_{\rm q}]$ the usual functional measure. Within the saddle-point approximation~\cite{kamenev_field_2011, altland_condensed_2010}, the tunneling rate is determined by the one instanton contribution~\cite{thompson_qubit_2022} and takes the form
\begin{align}\label{eq:Bit-flip_rate}
\begin{split}
    \Gamma_{i\to j} 
     &\propto e^{iS_{i\to j}}
\end{split}
\end{align}
where $S_{i \to j}$ is the action evaluated on the path which extremizes the action.

\iffalse 
Then, we perform a canonical transformation on the Keldysh field variables, the phase-space variable $a_{\rm cl}$ and the noise variable $a_{\rm q}$ to the new variables $b_{\rm cl}$, $b_{\rm q}$ (see~\cite{SM}). This transformation is chosen such that the Lindbladian density is quadratic in $b_{\rm q}$ and hence obtaining a Gaussian noise. The action is $S[b_{\rm cl}, b_{\rm q}] = \int dt \left(b_{\rm q}^*i \partial_t b_{\rm cl} -  b_{\rm q}i\partial_t b_{\rm cl}^* - i\mathcal{L}[b_{\rm cl}, b_{\rm q}]\right)$ where $\mathcal{L}$ is the Lindbladian density functional given in the \ref{SM}. The probability of tunneling from $\alpha_i$ to $\alpha_j$ for distinct $i, j \in \{1,2\}$ can be written as $\int_{
\substack{
b_{\rm cl}(- t/2) =  \alpha_{i} 
}
}^{b_{\rm cl}(t/2) =  \alpha_{j} }
\mathcal D[b_{\rm cl}, b_{\rm q}]
e^{i S[b_{\rm cl}, b_{\rm q}]}$
with $\mathcal{D}[b_{\rm cl}, b_{\rm q}]$ the usual functional measure. Within the saddle-point approximation  \cite{kamenev_field_2011, altland_condensed_2010}, the tunneling rate is determined by the one instanton contribution \cite{thompson_qubit_2022} and takes the form
\fi    

Varying the action with respect to the fields, we find that such action-extremizing paths must satisfy the equations of motion (EOM),
\begin{align}\label{eq:EOM_keldysh}
\begin{split}
    \partial_t{a}_{\rm cl} = &\partial_{a_{\rm q}^*} \mathcal{L},
    \hspace{0.5cm}
    \partial_t{a}_{\rm cl}^* = -\partial_{a_{\rm q}} \mathcal{L},
    \\
    \partial_t{a}_{\rm q} = &\partial_{a_{\rm cl}^*}\mathcal{L},
        \hspace{0.5cm}
    \partial_t{a}_{\rm q}^* = -\partial_{a_{\rm cl}} \mathcal{L}.
\end{split}
\end{align} 
These are a set of Hamilton's equations, where the phase-space and noise fields are conjugate variables $\{a_{\rm cl}, a_{\rm q}^*\} = \{a_{\rm q}, a_{\rm cl}^*\} = 1$, with all other Poisson brackets vanishing. The functional $\mathcal{L}$ generates the EOM, serving as an effective Hamiltonian, and is thus a conserved quantity, $\partial_t{\mathcal{L}}=0$. Our task is to the solve the set of coupled non-linear differential equations \cref{eq:EOM_keldysh} for the phase-space variable and the noise fields. These equations can be numerically integrated~\cite{lee_arealtime_2024}, and in some cases conserved quantities make the motion integrable~\cite{thompson_qubit_2022, thompson_population_2025}. In the following we provide an analytical solution using an Ansatz inspired by systems with classical detailed balance.  

Before presenting the construction of this Ansatz, we make some remarks on the form of solution. The EOM \cref{eq:EOM_keldysh} reduce to the mean-field equations \cref{eq:mean-field_eq} whenever the noise field is vanishing ${a_{\rm{q}}(t)=a^*_{\rm{q}}(t)=0}$. 
Momentarily ignoring boundary conditions, we refer to their common solution 
$[a_{\rm cl}(t), a_{\rm cl}^*(t) , a_{\rm q}(t), a_{\rm q}^*(t)] = [\alpha(t),\alpha^*(t), 0, 0]$ as the noise-free solution. Noise-free solutions alone cannot flow from one classically stable fixed point to another. That is, a switching path connecting the classically stable fixed points has to leave the noise-free subspace (see \cref{fig:drawing}). We look for solutions which start in the noise-free subspace ${a_{\rm q}=a_{\rm q}^*=0}$, thereby restricting our solutions to the $\mathcal{L} = 0$ subspace~\cite{SM}. Finally, we remark that the saddle-point equations \cref{eq:EOM_keldysh} do not necessarily respect complex conjugation, which implies that we have to deform the contour over which the path integral \cref{eq:Bit-flip_rate} is calculated, as done in the literature~\cite{kamenev_field_2011,ke_calculating_2025}. To highlight this, we rename the fields $a_{\rm cl,q}, a_{\rm cl,q}^* \to a_{\rm cl, q}, \overline{a}_{\rm cl,q} $ with $\overline{a}_{\rm cl,q}$ different from the complex conjugate of $a_{\rm cl,q}$.

\begin{figure}[t!]
    \centering
    \includegraphics[width=0.9\linewidth]{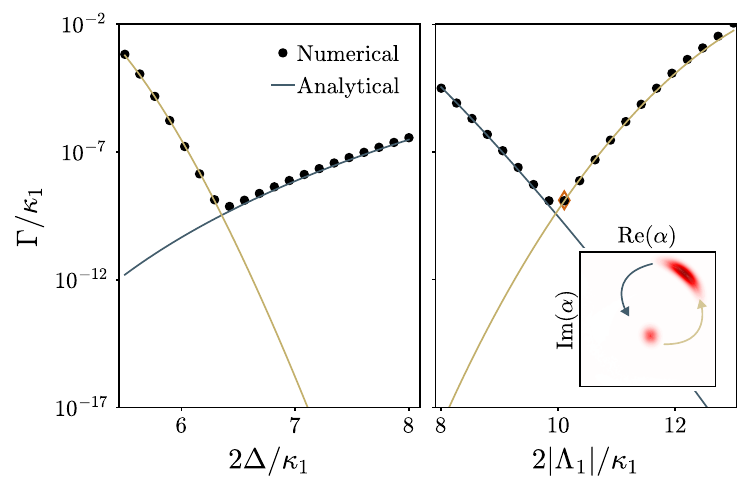}
     \caption{Dissipative gap of the driven dissipative Kerr oscillator: This figure reproduces a result obtained in~\cite{lee_arealtime_2024} by numerical integration of the EOM, with our analytical formula. The two solid lines represent the switching rate from the bright to the dim state and vice versa; the dissipative gap is the average of these switching rates~\cite{lee_arealtime_2024}. The dependence with respect to the detuning and the single-photon drive is shown on the left and right panel respectively. The inset represents the Wigner function of the steady-state at the position of the red diamond. Arrows indicate the direction of the switching rates between the fixed points in corresponding colors. Parameters $K/\kappa_1=0.1$, $2\Delta/\kappa_1= 6.33$, $2|\Lambda_1|\kappa_1=10$, noting the different definition of $\kappa_1$ in the present work. The prefactors to the predicted scaling \cref{eq:Bit-flip_rate} are the same in the left and right panel, $0.3$ and $0.05$ for the blue and gold curves, respectively.}
    \label{fig:Ginossar}
\end{figure}

\textit{Time-reversed paths}.--- To find the switching path, we first make a canonical transformation to a new coordinate system ${[a_{\rm cl}, \bar{a}_{\rm cl}, a_{\rm q}, \bar{a}_{\rm q}] \to [b_{\rm cl}, \bar{b}_{\rm cl}, b_{\rm q}, \bar{b}_{\rm q}]}$ defined by
\begin{align}\label{eq:b_q_def}
    \begin{split}
    b_{\rm cl} &= \frac{1}{\sqrt{2}}(a_{\rm cl} + a_{\rm q})
    ,
    \hspace{0.5cm}
    b_{\rm q} = \sqrt{2} a_{\rm q},
    \\
    \overline{b}_{\rm cl} &= \frac{1}{\sqrt{2}}(\overline{a}_{\rm cl} - \overline{a}_{\rm q})
    ,
    \hspace{0.45cm}
    \overline{b}_{\rm q} = \sqrt{2}\overline{a}_{\rm q},
\end{split}
\end{align}
which eliminates the terms in the Lagrangian density that are higher than second order in the noise fields $b_{\rm q}$ and $\bar{b}_{\rm q}$. Remarkably, once this coordinate transformation has been performed, the solution essentially mimics that used in classical systems with detailed balance~\cite{kamenev_field_2011,SM}.

Indeed, just as in the classical case, we find a parametrization of the noise fields in terms of the phase-space fields that satisfies the $\mathcal{L}=0$ condition~\cite{SM}
\begin{align}\label{eq:TR_parametrization}
\begin{split}
    {b}_{\rm q}
    &= 2{b}_{\rm cl} - \frac{2{K}_1^* \overline{b}_{\rm cl} + 2\Lambda_{3} \overline{b}_{\rm cl}^{2} + 2{\Lambda_{1}}^*}{{K}_2^* \overline{b}_{\rm cl}^{2} - 2 {\Lambda_{2}}^* - 2 \Lambda_{3}^* \overline{b}_{\rm cl}},\\
    \bar{b}_{\rm q} 
    &= -2\bar{b}_{\rm cl} + \frac{2{K_1}{b_{\rm cl}} + 2\Lambda_{3}^* {b_{\rm cl}}^{2} + 2\Lambda_{1}}{{K_2} {b_{\rm cl}}^{2} - 2 \Lambda_{2} - 2 {\Lambda_{3}} {b_{\rm cl}}}.
\end{split}
\end{align} 
Using this parametrization, given a noise-free solution $[\alpha(t),\alpha^*(t),0,0]$ we can build the so-called time-reversed solution~\cite{SM}, 
\begin{align} 
\begin{split}
\label{eq:trsol}
    \left[b_{\rm cl}, \overline{b}_{\rm cl}, b_{\rm q}, \overline{b}_{\rm q}\right] = \left[\alpha(-t), \alpha^*(-t), b_{\rm q}(b_{\rm cl}, \overline{b}_{\rm cl}),\overline{b}_{\rm q}(b_{\rm cl}, \overline{b}_{\rm cl}) \right]
\end{split}
\end{align}
The time-reversed solution allows us to construct the sought-for switching path between the two classically stable fixed points, as follows. 
On the manifold defined by \cref{eq:TR_parametrization}, the fixed points $\alpha_1$ and $\alpha_2$ are unstable (recall that they are stable in the noise-free manifold defined by the constraint $b_{\rm q} = \bar{b}_{\rm q}=0$). On this manifold, the fixed points $\alpha_{1}$ and $\alpha_2$ flow to the fixed point $\alpha_u$ which is stable along the subspace parametrized by \cref{eq:TR_parametrization}. This is the first part of the switching path. Afterwards, the trajectory switches to the noise-free subspace and stabilizes at the other fixed point, completing the second (and the last) part of the switching path. Corroborating these remarks, the full switching path between $\alpha_2$ and $\alpha_1$ is schematically depicted on~\cref{fig:drawing} in red. 

\begin{figure}[t!]
    \centering
    \includegraphics[width=0.9\linewidth]{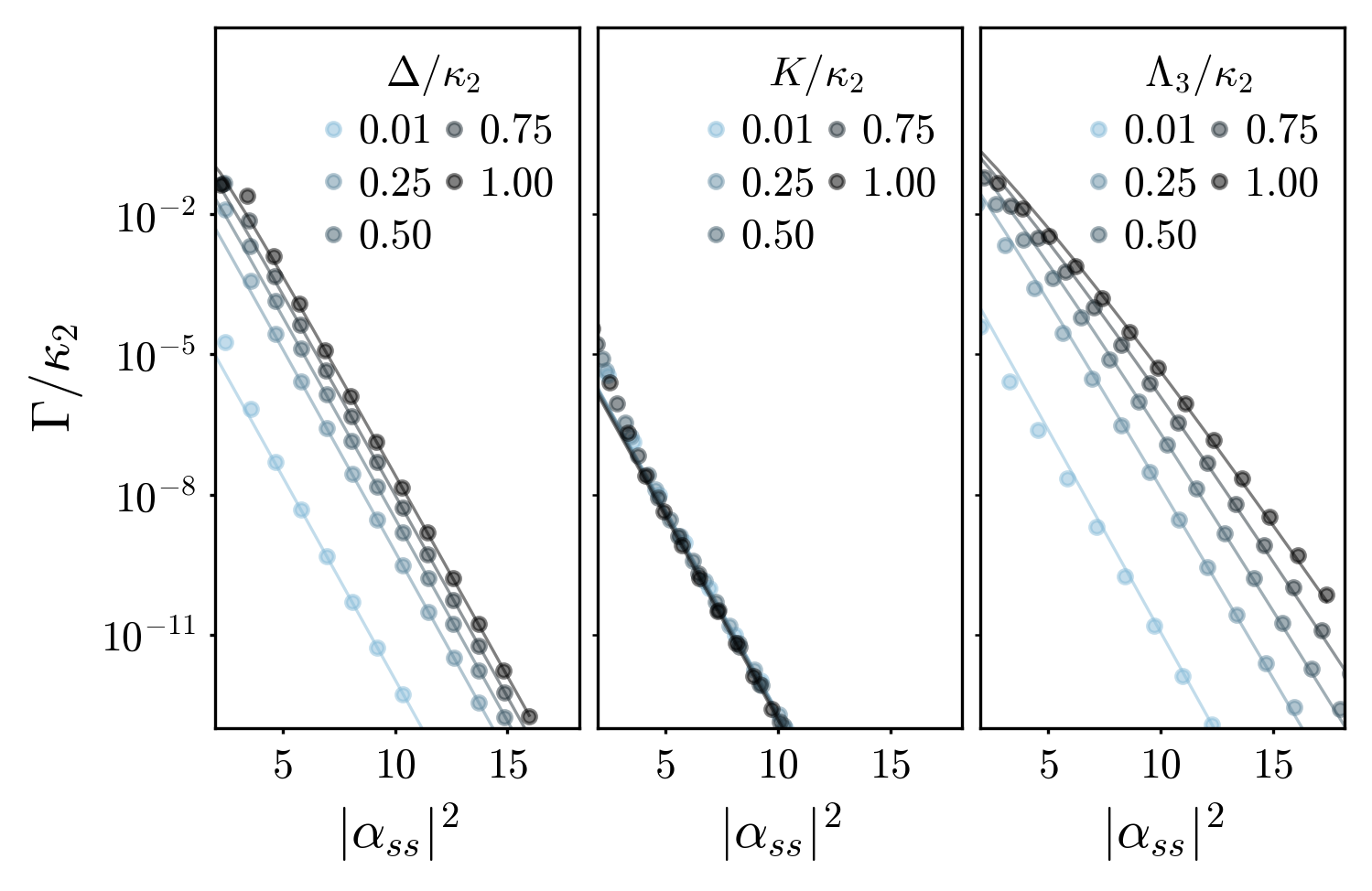}
    \caption{Bit-flip rate scaling for a dissipative cat-qubit: On each panel, the effect of a given imperfection on the scaling of the bit-flip rate is shown versus the steady-state amplitude. On the right, center and left panels, the colors indicate different values of the detuning, Kerr non-linearity and cubic photon drive imperfections respectively. The dots correspond to the smallest eigenvalue of the Lindbladian obtained via numerical diagonalization, whereas the lines show the predicted scaling. The prefactor for predicted scaling~\cref{eq:Bit-flip_rate} is set so that the prediction and numerical diagonalization cross at large $|\alpha_{ss}|^2$. In all panels, $ \kappa_1/\kappa_2 = 0.01$.}
    \label{fig:Bit-flip_cats}
\end{figure}

From the expression of the action $S[b_{\rm cl},b_{\rm q}]$ and since $\mathcal{L}=0$, the action accumulated on the noise-free subspace vanishes. Thus the first part of the switching path gives the only nonzero contribution to its action,
\begin{align}\label{eq:Action}
\begin{split}
    iS_{i\to j} &= \int 
    \left(
    -\overline{b}_{\rm q} \partial_t {b}_{\rm cl} + b_{\rm q} \partial_t \overline{b}_{\rm cl}
    \right)~\text{d}t\\
    &= \int 
    \left(-\overline{b}_{\rm q} \text{d} {b}_{\rm cl} + b_{\rm q} \text{d} \overline{b}_{\rm cl}
    \right)
    \\
    &= \left[\Phi(b_{\rm cl})\right]_{{b_{\rm{cl}}=\alpha_{i}}}^{{b_{\rm{cl}}=\alpha_{u}}}, 
\end{split}
\end{align} 
for distinct indices of the two classically stable fixed points $i,j \in \{1,2\}$. In the last line we used the parametrization \cref{eq:TR_parametrization}  and introduced a potential function~\cite{SM} that depends solely on the microscopic parameters of the Lindbladian $\mathbf{L}$ in \cref{eq:Lindblad_Eq}
\begin{align}
\begin{split}
    \Phi&(z) =2|z|^2 \\
    &+ 2\Re\left[-\frac{2\Lambda_3^*z}{K_2} + \frac{C_+}{K_2} \ln(z-z_+) + \frac{C_-}{K_2} \ln(z-z_-) \right],
\end{split}
\end{align}
where $z_\pm$ are the roots of $K_2 z^2 -2\Lambda_2 -2\Lambda_3z=0$ and $C_\pm = \frac{z_\pm (-2K_1K_2-4|\Lambda_3|^2) - 2\Lambda_1K_2-4\Lambda_3^*\Lambda_2}{K_2(z_\pm - z_\mp)}$.
By using the parametrization in \cref{eq:TR_parametrization} we have obtained that the integral does not depend on the trajectory. Remarkably, as shown in the \cite{SM}, $\Phi(z)$ is determined by the steady-state potential function of the complex-P representation, whose corresponding Fokker-Planck equation is known to obey detailed balance \cite{roberts_hidden_2021, BartoloPRA_2016}. This is analogous to Kramers theory~\cite{mylnikov_switching_2025}, which applies to systems obeying potential conditions.

Finally, we can obtain the switching rates using the saddle-point approximation \cref{eq:Bit-flip_rate}. We assume that the omitted prefactor -- referred to as the escape frequency~\cite{lee_arealtime_2024} and determined by the functional determinant that enters the saddle-point calculation\cite{altland_condensed_2010, kamenev_field_2011} -- is slowly varying with the system parameters. This prefactor can be calculated using Kramers theory~\cite{mylnikov_switching_2025}. Below, we confirm that hypothesis by comparing our analytical results to the numerical diagonalization of the Lindbladian. The dissipative gap is the average of the two rates $\Gamma = \frac{\Gamma_{1\to2} + \Gamma_{2\to1}}{2}$~\cite{lee_arealtime_2024}. In the limit where $\Lambda_2$ is the largest energy scale, which corresponds to cat-state stabilization, the two rates are equal~\cite{thompson_qubit_2022}. Furthermore, in this limit, the potential function reduces to $2|z|^2$. The explicit analytical solution to the saddle-point equations of motion via the time-reversed path, and the corresponding rate, \cref{eq:Action} and \cref{eq:Bit-flip_rate}, are the main results of this work. 

\begin{figure}[t!]
    \centering
    \includegraphics[width=\linewidth]{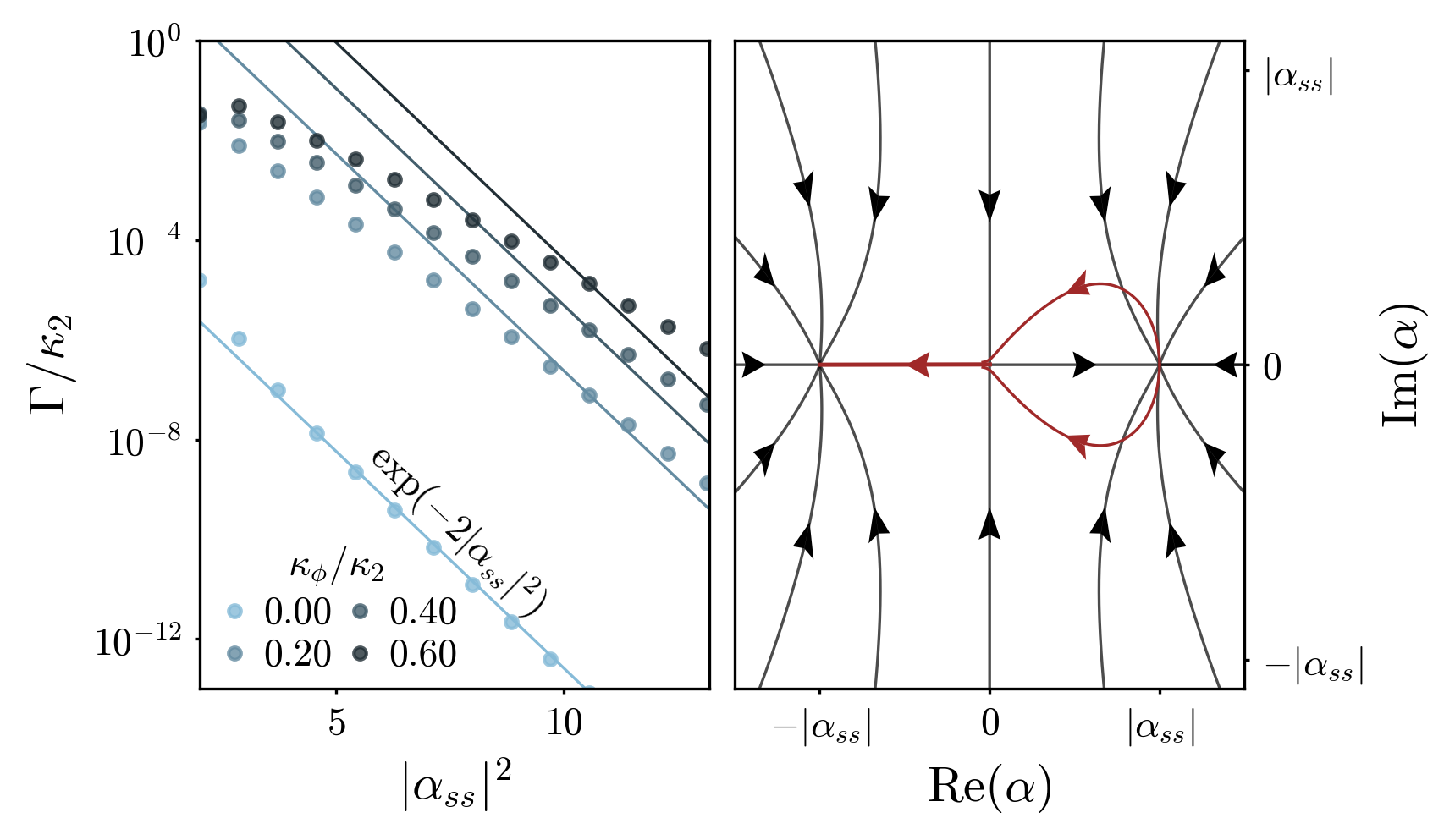}
    \caption{Study of the dephasing imperfection and breaking of the detailed balance condition. Left panel: bit-flip rate versus the steady-state amplitude for a dissipative cat-qubit for various values of a dephasing imperfection. The dots correspond to smallest eigenvalue of the Lindbladian obtained via numerical diagonalization. The lines show scaling in $e^{-2|\alpha_{ss}|^2}$. In contrast to systems with HTRS, we observe that a dephasing imperfection results in a scaling of the bit-flip rate that cannot be recast as $e^{-2|\alpha_{ss}|^2}$, even in the large $|\alpha_{\textit{ss}}|$ limit. As in \cref{fig:Bit-flip_cats}, we include a small single-photon imperfection $\kappa_1/\kappa_2=0.01$. Right panel: phase space flow and switching paths for a dissipative Kerr cat with a dephasing imperfection for $\kappa_1=0$ and $\kappa_\phi/\kappa_2 =0.4 $. In red, two noise-assisted switching paths obtained from the numerical integration of \cref{eq:EOM_keldysh} with initial conditions near the fixed-point $\alpha_{ss}$. The obtained trajectories are not the time-reversed partner of the noise-free  evolution (black line) from $0$ to $\alpha_{ss}$.  }
    \label{fig:dephasing_pert}
\end{figure}

\textit{Numerical investigation}.--- In \cref{fig:Ginossar}, we compare our results to a recent numerical investigation of the dissipative gap of a driven dissipative Kerr oscillator \cite{lee_arealtime_2024}. We numerically diagonalize the Lindbladian where $\kappa_2=\Lambda_2=\Lambda_3=0$ and look for the smallest eigenvalue when sweeping detuning and single-photon drive. The predicted scaling agrees with our numerical diagonalization, and with Fig.~3 of \cite{lee_arealtime_2024}. We choose the prefactor omitted in \cref{eq:Bit-flip_rate} to match the numerical data. This prefactor is the same in both panels of \cref{fig:Ginossar}, indicating a weak dependence on detuning and single-photon drive in this parameter regime.

In~\cref{fig:Bit-flip_cats}, we focus on the dissipative gap of bistable systems stabilizing cat states, which coincides with the bit-flip rate. As discussed in the SM, in the limit where $\Lambda_2$ is the largest energy scale, the fixed points have opposite signs $\alpha_1 = -\alpha_2$ and we denote the common amplitude by $|\alpha_{1,2}| = |\alpha_{ss}|$. In~\cref{fig:Bit-flip_cats}, the numerical diagonalization of the bit-flip rate has the same scaling as the prediction for varying detuning, Kerr and cubic photon-drive imperfections. As in the case of \cref{fig:Ginossar} discussed above, we further find that the prefactor of~\cref{eq:Bit-flip_rate} cannot exponentially depend on the parameters of the microscopic model \cref{eq:Lindblad_Eq}. Importantly, in each case, in the limit of large $\Lambda_2$, the bit-flip rate obtained from~\cref{eq:Bit-flip_rate} simplifies to $e^{-2|\alpha_{ss}|^2}$. Note that in general $\alpha_{ss}^2 \neq \alpha_0^2 = 2 \Lambda_2/\kappa_2$, the coherent state amplitude that enters the dissipator of an idealized dissipative cat qubit $\kappa_2\mathbf{D}_{\hat{a}^2 - \alpha_0^2}\hat{\rho}$: the rate we obtain is non-perturbative in all the microscopic parameters of the Lindbladian.

In the SM~\cite{SM}, we study the multi-mode bosonic system satisfying HTRS described in \cite{roberts_competition_2023}. For this multi-mode system, we explicitly calculate the potential function and extremize the action, following the procedure in \cref{eq:Action}. To compare with numerical simulations, we then focus on the two-mode case and identify a parameter regime that exhibits bistability. In this regime, the switching probability is calculated and subsequently used to estimate the dissipative gap. We report an agreement between the analytical calculations and the numerical computation of the dissipative gap.

\textit{Breaking the time-reversal symmetry}.--- So far we have considered imperfections that satisfy the detailed balance condition for the complex-P function, or equivalently obey HTRS~\cite{roberts_hidden_2021,SM}. Using this symmetry we obtained an extremal time-reversed path to calculate the switching rate. In~\cref{fig:dephasing_pert} we add a dephasing imperfection $\kappa_{\phi} \mathbf{D}_{\ha^\dagger\ha}$ to the Lindbladian of \cref{eq:Lindblad_Eq} in which we only keep $\kappa_2$ and $\Lambda_2$. This imperfection breaks the detailed balance condition discussed in \cite{SM}, and we can no longer find a time-reversed extremal path, equivalent to the inapplicability of Kramers theory~\cite{mylnikov_switching_2025}. We demonstrate this property in the right panel. Using the numerical integration of the EOM we obtain two trajectories taking the system from one of the stable fixed-point to the other~\cite{SM}. This is a direct confirmation that the time-reversal symmetry is broken, hence we cannot expect the method presented above to work in that case. Moreover, we observe a scaling that cannot be recast as $e^{-2|\alpha_{\textit{ss}}|^2}$ as highlighted by the left panel. 

\textit{Conclusion}.--- We have identified a set of single-mode bosonic systems in which we have predicted the dissipative gap. The latter we have evaluated from a saddle-point approximation of the Keldysh path integral by using a time-reversed Ansatz to analytically solve the corresponding saddle-point EOM. Moving beyond a single mode, our analysis suggests a path forward for obtaining dissipative gaps of many-body bosonic systems satisfying HTRS, which fall beyond current numerical possibilities. To that end, we have computed the dissipative gap of a specific many-body model in the SM \cite{SM}, and the methodology presented there can be extended to other models with HTRS. In the particular case of bosonic cat codes, understanding how the scaling of bit-flip rates is altered under experimental conditions is key to develop viable bosonic codes, given that such systems are prone to a variety of imperfections inducing additional decay mechanisms~\cite{carde_flux_2025}. More generally, this framework not only offers insights into open quantum systems but also paves the way for extending these results to generic models with HTRS~\cite{Yao_PRL_2025, lingenfelter_exact_2024, Roberts_PRL_2023_bosons, Roberts_PRL_2023}.

\textit{Acknowledgments}.--  The authors acknowledge fruitful discussions with Pierre Rouchon and Aashish Clerk. AM acknowledges funding from NSERC and the Canada First Research Excellence Fund.

\bibliographystyle{apsrev4-2}
\bibliography{ref}

\end{document}

\author{L{\ifmmode\acute{e}\else\'{e}\fi}on Carde}
\thanks{leon.carde@alice-bob.com}

\affiliation{Laboratoire de Physique de l'\'{E}cole Normale Sup\'{e}rieure$,$ Inria$,$ ENS$,$ Mines ParisTech$,$ Universit\'{e} PSL$,$ Sorbonne Universit\'{e}, Paris$,$ France}
\affiliation{Alice and Bob$,$ 53 Bd du G\'{e}n\'{e}ral Martial Valin$,$ 75015$,$ Paris$,$ France}
\author{Ronan Gautier}
\affiliation{Alice and Bob$,$ 53 Bd du G\'{e}n\'{e}ral Martial Valin$,$ 75015$,$ Paris$,$ France}
\author{Nicolas Didier}
\affiliation{Alice and Bob$,$ 53 Bd du G\'{e}n\'{e}ral Martial Valin$,$ 75015$,$ Paris$,$ France}
\author{Alexandru Petrescu}
\affiliation{Laboratoire de Physique de l'\'{E}cole Normale Sup\'{e}rieure$,$ Inria$,$ ENS$,$ Mines ParisTech$,$ Universit\'{e} PSL$,$ Sorbonne Universit\'{e}, Paris$,$ France}
\author{Joachim Cohen}
\affiliation{Alice and Bob$,$ 53 Bd du G\'{e}n\'{e}ral Martial Valin$,$ 75015$,$ Paris$,$ France}
\author{Alexander McDonald}
\affiliation{Institut Quantique and D\'{e}partement de Physique\protect\\
Universit\'{e} de Sherbrooke, Sherbrooke~J1K~2R1~Qu\'{e}bec, Canada}

\date{\today}

\title{SUPPLEMENTARY MATERIAL\\
Non-perturbative switching rates in bistable open quantum systems:\\
from driven Kerr oscillators to dissipative cat qubits}
\maketitle

This Supplementary Material is organized as follows. In \cref{sec:ld} we derive the Lindbladian density for the model considered in the main text.  In \cref{app:sec:fixed-points}, we analyze the fixed points of the mean-field equation. Then in \cref{app:sec:Time-R}, we perform a canonical transformation so that the Lindbladian density becomes quadratic in the noise fields. We then make an analogy with the detailed balance condition for the complex-P function and define the time-reversed Ansatz used in the main text [MT]. We find a condition under which this Ansatz is valid and compare it with the detailed balance condition of the complex-P function and equivalently HTRS. In \cref{app:sec:Action}, we explicitly calculate the extrema of the action and derive the potential function derived in the main text. Moreover we calculate the potential of the steady-state complex-P function and show a connection between both quantities. In \cref{app:sec:numint} we provide the details of the numerical integration of the equations of motion if dephasing enters the Lindbladian.

\section{Lindbladian density of a single-mode HTRS system}
\label{sec:ld}
In this section we derive the Lindbladian density of the system analyzed in the main text, by standard methods~\cite{sieberer_keldysh_2016, kamenev_field_2011}. We consider a single bosonic mode with hidden time-reversal symmetry (HTRS)~\cite{roberts_driven_2020} described by the Lindblad master equation in the main text. The corresponding Keldysh action functional ~\cite{sieberer_keldysh_2016, kamenev_field_2011} is 
\begin{align}\label{app:eq:Action_Keldysh}
S[a_{\rm cl}, a_{\rm q}] = \int dt \left(a_{\rm q}^*i \partial_t a_{\rm cl} -  a_{\rm q}i\partial_t a_{\rm cl}^* - i\mathcal{L}[a_{\rm cl}, a_{\rm q}]\right),
\end{align}
where the Lindbladian density functional $\mathcal{L}$  is 
\begin{widetext}
\begin{align}\label{app:eq:Lindbladian_density}
\begin{split}
    \mathcal{L}[a_{\rm cl}, a_{\rm q}] =& 
    \Bigg\{a_{\rm q} \left[a_{\rm cl}^*\left(-iK_1^* + \frac{iK_2^*}{2} |a_{\rm cl}|^2 \right) - \frac{ i \Lambda_{3} a_{\rm cl}^{*2}}{\sqrt{2}}- \sqrt{2} i \Lambda^{*}_{1} - 2 i \Lambda^{*}_{2} a_{\rm cl} - \sqrt{2} i \Lambda^{*}_{3} |a_{\rm cl}|^2 \right]\\
    &+ \frac{a_{\rm q}^{2}a_{\rm q}^*}{2} \left[i K_2 a_{\rm cl}^* - \sqrt{2} i \Lambda^{*}_{3}  \right] - \text{c.c} \Bigg\} + a_{\rm q} a_{\rm q}^* \left(- 2 |a_{\rm cl}|^2 \kappa_{2} - \kappa_{1}\right).
\end{split}
\end{align}
\end{widetext}
The field $a_{\rm cl}$ and its complex conjugate represent the phase-space fields, while $a_{\rm q}$ and its conjugate are the noise fields. 

\section{Fixed points of a single-mode HTRS system}\label{app:sec:fixed-points}
In this appendix we detail the calculation of the fixed points of a single-mode system satisfying HTRS. 
Our starting point is the mean-field equation for the expectation value of the annihilation operator $ \langle\ha(t)\rangle = \alpha(t)$, which reads
\begin{equation}\label{eq:mean-field_eq}
\begin{split}
    i\partial_t \alpha &= \alpha K_1 - \alpha |\alpha|^2 K_2 +\Lambda_{1} + 2\alpha^*\Lambda_2 +\alpha^2 \Lambda_3^* + 2 |\alpha|^2 \Lambda_3
\end{split}
\end{equation}
The fixed points $z$ are static solutions of the mean-field equation. After dividing by $K_2$ and setting the time derivative to $0$ in the mean-field \cref{eq:mean-field_eq}, we obtain a polynomial equation for the fixed points
\begin{align}\label{app:eq:fixed-point}
\begin{split}
    |z|^2z  - \frac{2\Lambda_3}{K_2} |z|^2 -\frac{\Lambda_3^*}{K_2} z^2 - \frac{K_1}{K_2} z -\frac{2\Lambda_2}{K_2} z^* - \frac{\Lambda_1}{K_2}=0.
\end{split}
\end{align}
Note that the same equation is recovered from the equations of motion obtained by extremizing the Keldysh action, in the noise-free subspace where $a_{\rm q}=0$. We now solve this equation in a several cases. 

\textit{Cat-state stabilization.} Considering the limit $\frac{\Lambda_2}{K_2}\gg1$, as done in the main text for this case, we rescale ${z = w/\epsilon}$, where $\epsilon = \left|\frac{K_2}{2\Lambda_2}\right|^{1/2} \ll 1$, and rewrite the equation for the fixed points \cref{app:eq:fixed-point} as
\begin{align}
    \begin{split}
        w^*\left(w^2-e^{i\psi_2} \right) - \epsilon \frac{2\Lambda_3}{K_2} |w|^2- \epsilon\frac{\Lambda_3^*}{K_2} w^2&  \\
        - \epsilon^2\frac{K_1}{K_2} w - \epsilon^3 \frac{\Lambda_1}{K_2}&=0
        \end{split}
\end{align}
where $\psi_2 = \arg\frac{2\Lambda_2}{K_2}$. We now expand $w = w_0 + \epsilon w_1$ and solving the above equation order-by-order in $\epsilon$, to obtain three solutions. The first is $w_0 =w_1=0$, corresponding to the unstable fixed point, whereas the two stable fixed points are given by
\begin{align}
    \begin{split}
        w_0 = \pm e^{i\psi_2/2}, \; w_1 = \frac{2\Lambda_3 + \Lambda_3^*e^{i\psi_2}}{2K_2}.
    \end{split}
\end{align}
Summarizing the above we have one trivial solution $z = O\left(\epsilon\right)$ and two solutions corresponding to the stable fixed points used for the encoding of the cat state,
\begin{align}
    \begin{split}
        z = \pm \left|\frac{2\Lambda_2}{K_2}\right|^{1/2} e^{i\psi_2/2} + \frac{2\Lambda_3 + \Lambda_3^*e^{i\psi_2}}{2K_2} + O\left(\epsilon\right).
    \end{split}
\end{align}

\textit{Driven dissipative Kerr oscillator.} For the driven dissipative Kerr oscillator we have $\Lambda_3 =\Lambda_2 = \kappa_2= 0$, and the fixed point equation \cref{app:eq:fixed-point} reads
\begin{align}
\label{app:eq:fpKO}
    |z|^2z - \frac{K_1}{K_2} z - \frac{\Lambda_1}{K_2}=0
\end{align}
Denoting $|z|^2 = r^2$, we recast this equation to $z=\frac{\Lambda_1}{r^2K_2 -K_1 }$. Therefore we have one fixed point per non-negative value of $r$. Injecting $z$ into the equation ${|z|^2 = r^2}$ we obtain an equation for $R = r^2$,
\begin{align}
    R^3 |K_2|^2 -2\Re(K_1^*K_2) R^2 + |K_1|^2 R -|\Lambda_1|^2= 0.
\end{align}
We have a third order equation that can have up to 3 roots, each root corresponding to a unique fixed point. Indeed, each root $R$ corresponds to a unique $r>0$, which in turn corresponds to a unique fixed point $z=\frac{\Lambda_1}{RK_2 -K_1 }$.

To solve this third order equation we define $U = R -\frac{2 \mathrm{Re}(K_1^*K_2)}{3|K_2|^2}$, and we obtain a depressed cubic equation $U^3 + pU+q =0$ with,
\begin{align}
    \begin{split}
        p &= \frac{|K_1|^2}{|K_2|^2} -\frac{4\Re(K_1^*K_2)^2}{3|K_2|^4},\\
        q &= -2\left(\frac{2\Re(K_1^*K_2)}{3|K_2|^2}\right)^3 - \frac{|\Lambda_1|^2}{|K_2|^2}- \frac{2\Re(K_1^*K_2)|K_1|^2}{3|K_2|^4}.
    \end{split}
\end{align}
The resulting cubic has three real roots if and only if the discriminant is negative, $4p^3 + 27q^2<0$. The discriminant reads,
\begin{align}
    \begin{split}
        &\left|\frac{K_1}{K_2}\right|^6 \left[2^7\cos(\phi)^4/3 + 4\sin^2(\phi)\right]\\
        &+\left|\frac{K_1}{K_2}\right|^3 \left|\frac{\Lambda_1}{K_2}\right|^2 \left[2^5\cos(\phi)^2 + 2^23^2\right] \cos(\phi)\\
        &+\left|\frac{\Lambda_1}{K_2}\right|^4 27,
    \end{split}
\end{align}
where we have defined $ \cos(\phi) |K_1K_2| = \Re(K_1^*K_2)$. We can rearrange the condition to have three real roots to the following form
\begin{align}
    0<\frac{|K_1|^3}{|K_2||\Lambda_1|^2} P[\cos(\phi)]
    +\frac{|K_2||\Lambda_1|^2}{|K_1|^3} Q[\cos(\phi)]
    <-\cos(\phi) ,
\end{align}
where $P[\cos(\phi)]>0 $ and $Q[\cos(\phi)]>0 $ are rational functions of $\cos(\phi)^2$.

The region of bistability of the driven dissipative Kerr oscillator (with two photon dissipation) is such that $|K_1|^3$ ${\approx}$ $|K_2||\Lambda_1|^2$ 
and such that $\cos(\phi)<0$. 
When the bistability condition is met we have three real roots
\begin{align}
\begin{split}
    R_k =& \frac{2\Re(K_1^*K_2)}{3|K_2|^2} \\
    &+ 2\sqrt{\frac{-p}{3}}\cos\left[ \frac{1}{3} \arccos\left( \frac{3q}{2p}\sqrt{\frac{-3}{p}}\right) +\frac{2\pi k}{3}\right],
\end{split}
\end{align}
for $k=0,1,2$. To obtain the predictions in the main text we solved for the fixed-point equation \cref{app:eq:fixed-point} numerically, instead of using the above solutions. The numerical cost of finding the roots of a set of polynomial equations is constant with the value of the parameters and does not limit the prediction of the dissipative gap.

\section{Canonical transformation and time-reversed Ansatz}\label{app:sec:Time-R}

As explained in the main text, we  perform an analytic continuation of the action \cref{app:eq:Action_Keldysh} and rename $a_{\rm cl/q}^*$ to $\overline{a}_{\rm cl/q}$ to highlight that we can modify the contour on which the action is integrated and that $\overline{a}_{\rm cl/q}$ are no longer the complex conjugates of $a_{\rm cl/q}$ \cite{kamenev_field_2011}.  To find the time-reversed path in the main text, we perform a canonical transformation ${[a_{\rm cl}, \bar{a}_{\rm cl}, a_{\rm q}, \bar{a}_{\rm q}] \to [b_{\rm cl}, \bar{b}_{\rm cl}, b_{\rm q}, \bar{b}_{\rm q}]}$ to eliminate the terms in the Keldysh action that are third order or higher in the noise field 
\begin{align}\label{app:eq:canonical_transformation}
\begin{split}
    b_{\rm cl} &= \frac{1}{\sqrt{2}}(a_{\rm cl} + a_{\rm q})
    ,
    \hspace{0.5cm}
    b_{\rm q} = \sqrt{2} a_{\rm q},
    \\
    \overline{b}_{\rm cl} &= \frac{1}{\sqrt{2}}(\overline{a}_{\rm cl} - \overline{a}_{\rm q})
    ,
    \hspace{0.45cm}
    \overline{b}_{\rm q} = \sqrt{2}\overline{a}_{\rm q},
\end{split}
\end{align}
which preserves the Poisson brackets $\left\{b_{\rm cl},\bar{b}_{\rm q} \right\} = \left\{\bar{b}_{\rm cl},-b_{\rm q} \right\} = 1$, and turns the Lindbladian density \cref{app:eq:Action_Keldysh} into
\begin{align}
    \begin{split}
    \mathcal{L} =  \begin{bmatrix}
            \overline{b}_{\rm q} & -b_{\rm q}
        \end{bmatrix}\left(\mathbb{A} + \mathbb{D}\begin{bmatrix}
            \overline{b}_{\rm q}\\
            -b_{\rm q}
        \end{bmatrix}\right) \equiv \mathbb{P}_i \mathbb{A}_i + \mathbb{P}_i \mathbb{P}_j \mathbb{D}_{ij},
        \label{app:eq:fw}
    \end{split}
\end{align}
where after the second equality we used Einstein summation over the indices $i,j \in \{1,2\}$, and have introduced complex 2-vectors of the momenta $\mathbb{P}^T \equiv \left[\bar{b}_q, -b_q \right] \equiv \left[ \mathbb{P}_1, \mathbb{P}_2 \right]$ and coordinates $\mathbb{X}^T = \left[ b_{cl}, \bar{b}_{cl} \right] \equiv \left[ \mathbb{X}_1, \mathbb{X}_2 \right]$, obeying Poisson brackets $\{ \mathbb{X}_i , \mathbb{P}_j \} = \delta_{ij}$. Additionally we denoted $\mathbb{A}^T = \left[
    A, \overline{A}
\right]$, a complex 2-vector, and $\mathbb{D}= \text{diag}\left[ D, \overline{D} \right]$ a complex-valued diagonal $2 \times 2$ matrix, expressed in terms of four entire functions of the classical complex coordinates $\mathbb{X}_{1,2}$ only
\begin{align}\label{app:eq:HTRS_AD}
    \begin{split}
        \overline{D} &= -i\frac{K_2^*}{2}\overline{b}_{\rm cl}^2 + i\Lambda_2^* + i \Lambda_3^*\overline{b}_{\rm cl},\\
        {D} &= i\frac{K_2}{2}{b}_{\rm cl}^2 - i\Lambda_2 - i \Lambda_3{b}_{\rm cl},\\
        \overline{A}&= 2b_{cl}\overline{D} + iK_1^*\overline{b}_{\rm cl} + i\Lambda_3\overline{b}_{\rm cl}^2+i\Lambda_1^*,\\
        {A}&= 2\overline{b}_{cl}D - iK_1{b}_{\rm cl} - i\Lambda_3^
        *{b}_{\rm cl}^2-i\Lambda_1.
    \end{split}
\end{align}
$\mathcal{L}$ of \cref{app:eq:fw} is itself an entire function of the complex momenta $\mathbb{P}_{1,2}$ and coordinates $\mathbb{X}_{1,2}$. As it contains terms up to quadratic in the noise fields $\mathbb{P}_i$, $\mathcal{L}$ is a complex-valued generalization of the Freidlin-Wentzell Hamiltonian \cite{pinna_large_2016,Freidlin2012}, with $\mathbb{A}$ corresponding to the drift field and $\mathbb{D}$ the diffusion tensor of the corresponding Fokker-Planck equation derived below.

\textit{The time-reversed Ansatz}. The equations of motion deriving from the Lindbladian density $\mathcal{L}$ of \cref{app:eq:fw} take the form of Hamilton equations $\dot{\mathbb{X}}_i = \frac{\partial \mathcal{L}}{\partial \mathbb{P}_i}$ and $\dot{\mathbb{P}}_i = -\frac{\partial \mathcal{L}}{\partial \mathbb{X}_i}$, or explicitly
\begin{align}
    \begin{split}
    \dot{\mathbb{X}}_i &= \mathbb{A}_i + 2 \mathbb{P}_j \mathbb{D}_{i j}, \\
    \dot{\mathbb{P}}_i &= - \mathbb{P}_j \partial_i \mathbb{A}_j - \mathbb{P}_j \mathbb{P}_k \partial_i \mathbb{D}_{jk}, \label{ap:eq:om}
    \end{split}
\end{align}
where in the second line we have used $\mathbb{D}_{ij}=\mathbb{D}_{ji}$, and we use shorthand $\partial_i = \partial/\partial \mathbb{X}_i$. Following \cite{kamenev_field_2011}, we can find a noise-free solution to the equations of motion \cref{ap:eq:om}, where the velocities of the classical coordinates follow the drift field
\begin{align}
    \begin{split}\label{ap:eq:rel}
        \dot{\mathbb{X}}_i &= \mathbb{A}_i,\\
        \mathbb{P}_i &= 0,
    \end{split}        
\end{align}
which will correspond to a relaxation path. 

To describe the switching path, note that $\mathcal{L}(\mathbb{P},\mathbb{X})=0$ for $\mathbb{P}=0$ and $\mathcal{L}$ is conserved, so the sought-for switching path will belong to the same manifold, that is  
\begin{align}
    \mathcal{L} = \mathbb{P}_i \mathbb{A}_i + \mathbb{P}_i \mathbb{P}_j \mathbb{D}_{ij} = 0.
\end{align}
Noting the implicit sum, we construct an Ansatz for the solution to \cref{ap:eq:om} by forcing each term of the sum to vanish, that is, for each $i=1,2$
\begin{align}
    \mathbb{A}_i +  \mathbb{P}_j \mathbb{D}_{ij} = 0, 
\end{align}
or equivalently, if $\mathbb{D}$ is invertible at all $\mathbb{X}$, $\mathbb{P} = - \mathbb{D}^{-1} \mathbb{A}$. Plugging this Ansatz into \cref{ap:eq:om} we find that the velocities corresponding to the classical coordinates have to follow the negative of the drift field, \ie they must necessarily obey a time-reversed equation with respect to \cref{ap:eq:rel}
\begin{align}
    \begin{split}\label{ap:eq:swi}
        \dot{\mathbb{X}}_i &= -\mathbb{A}_i,\\
        \mathbb{P}_i &= - \mathbb{D}^{-1}_{ij} \mathbb{A}_j
    \end{split}    
\end{align}
Conversely, assuming that the velocities of the coordinates obey the time-reversed equation $\dot{\mathbb{X}}_i = -\mathbb{A}_i$ together with the first equation of motion \cref{ap:eq:om} implies $\mathbb{P}_i = - \mathbb{D}^{-1}_{ij} \mathbb{A}_j$. Having proven this equivalence, we will further refer to \cref{ap:eq:swi} as the (unique) \textit{time-reversed Ansatz}, and to the second \cref{ap:eq:swi} as the \textit{time-reversed parametrization}.

We now derive a necessary and sufficient condition for the time-reversed parametrization in \cref{ap:eq:swi} to obey the equations of motion \cref{ap:eq:om}. This is equivalent checking the conditions under which $\dot{\mathbb{P}}_i = - \partial_i \mathcal{L}$. By \cref{ap:eq:swi} the left-hand side of this equation rewrites to
\begin{align}
\begin{split}
    \dot{\mathbb{P}}_i =& - \dot{\mathbb{D}}_{ij}^{-1} \mathbb{A}_j - \mathbb{D}_{ij}^{-1} \dot{\mathbb{A}}_j = - \partial_k \mathbb{D}_{ij}^{-1} \dot{\mathbb{X}}_k \mathbb{A}_j - \mathbb{D}_{ij}^{-1} \partial_k \mathbb{\mathbb{A}}_j \mathbb{X}_k \\
    =& (\mathbb{A}_k \partial_k) ( \mathbb{D}_{ij}^{-1} \mathbb{A}_j).
    \end{split}
\end{align}
where, in the first line, we used the chain rule to express the total time derivative in terms of partial derivatives with respect to position and total derivatives of position, and to pass to the second line we used the time-reversed equation of motion for position, the first \cref{ap:eq:swi}. The right-hand-side of the Hamilton equation of motion for $\mathbb{P}_i$ reads
\begin{align}
\begin{split}
    -\partial_i \mathcal{L} =& -\mathbb{P}_k \partial_i \mathbb{A}_k - \mathbb{P}_k \mathbb{P}_l \partial_i \mathbb{D}_{kl} \\
    =& \mathbb{D}_{kl}^{-1} \mathbb{A}_l \partial_i \mathbb{A}_k - \mathbb{D}_{km}^{-1} \mathbb{A}_m  \mathbb{D}_{ln}^{-1} \mathbb{A}_n \partial_i \mathbb{D}_{kl} \\
    =& \mathbb{D}_{kl}^{-1} \mathbb{A}_l \partial_i \mathbb{A}_k  + \partial_i \mathbb{D}_{kl}^{-1} \mathbb{A}_k \mathbb{A}_l \\
    =& \mathbb{A}_l \partial_i (\mathbb{D}_{kl}^{-1} \mathbb{A}_k),
    \end{split}
\end{align}
where to obtain the second line, we have plugged in the time-reversed parametrization, second \cref{ap:eq:swi}, and, to obtain the third line, we used $\partial_i \mathbb{D}_{kl} \mathbb{D}_{ln}^{-1}=0$.

Thus the time-reversed parametrization \cref{ap:eq:swi} obeys the equations of motion \cref{ap:eq:om} if and only if the following hold for $i=1,2$
\begin{align}
    \mathbb{A}_l \left[  \partial_l (\mathbb{D}^{-1}_{ij} \mathbb{A}_j) - \partial_i (\mathbb{D}_{kl}^{-1} \mathbb{A}_k) \right] = 0
\end{align}
Assuming ${A}_l(\mathbb{X}) \neq 0$, which is valid away from the fixed points, the equation above is equivalent to $ \partial_i (\mathbb{D}_{jl}^{-1}\mathbb{A}_l ) = \partial_j (\mathbb{D}_{il}^{-1}\mathbb{A}_l )$ which we write concisely as
\begin{align}
\label{ap:eq:smallDB}
        \partial_j Z'_i = \partial_i Z'_j,\; Z'_i \equiv  -\mathbb{D}_{i j}^{-1} \mathbb{A}_{j},
\end{align}
\ie the vector field defined by the time-reversed parametrization in \cref{ap:eq:swi} is curl-free. This is to be contrasted to the usual potential condition \cite{gardiner_handbook_2009, walls_quantum_2025}, \cref{ap:eq:DB} below. To summarize, the time-reversed parametrization is a solution to the equations of motion if and only if it is curl-free in the sense of \cref{ap:eq:smallDB}.

\section{Fokker-Planck equation and potential condition}
The Lindbladian analyzed in the main text, which obeys HTRS, equivalently should correspond to a Fokker-Planck equation for the complex-P function that obeys detailed balance ~\cite{BartoloPRA_2016, roberts_hidden_2021}. To make the detailed balance condition explicit here, following Drummond and Gardiner~\cite{drummond_generalised_1980}, we start from the Lindblad master equation $\partial_t \hat{\rho} = \mathbf{L}(\hat{\rho})$ and obtain a partial differential equation for the complex-P function $P(\alpha,\beta)$, defined in relation to the density matrix $\hat{\rho}$ as
\begin{align}
    \hat{\rho} = \int_C \int_{C'} \frac{\ket{\alpha}\bra{\beta^*}}{\braket{\beta^*}{\alpha}} P(\alpha,\beta) d\alpha d\beta,
\end{align}
where the contours of integration $C,C'$ are closed contours whose definition depends on $\hat{\rho}$~\cite{BartoloPRA_2016,drummond_generalised_1980}.
To this aim, we use the following rules to convert the Lindbladian evolution of $\hat{\rho}$ into an equation for the time evolution of the complex-P function~\cite{drummond_generalised_1980}
\begin{align}
    \begin{split}
        \hat{a}\hat{\rho} &\to \alpha P,\\
        \hat{a}^\dagger\hat{\rho} &\to (\beta - \partial_\alpha)P,\\
        \hat{\rho}\hat{a} &\to (-\partial_{\beta} + \alpha) P,\\
        \hat{\rho}\hat{a}^\dagger &\to \beta P,
    \end{split}
\end{align}
where we have used shorthand $\partial_\alpha = \partial/\partial \alpha$, $\partial_\beta = \partial/\partial \beta$. 

Applying these rules we obtain a Fokker-Planck partial differential equation for $P$
\begin{align}
\begin{split}
    \partial_t P =& \partial_{\alpha}\left[-A(\alpha,\beta) + \partial_\alpha D(\alpha)\right] P \\
        & + \partial_{\beta}\left[-\overline{A}(\alpha,\beta) + \partial_{\beta}\overline{D}(\beta)\right] P,
\end{split}
\end{align}
with the understanding that the partial derivatives act on all elements to their right.
Upon identifying $(\alpha,\beta) \equiv \mathbb{X} = (\mathbb{X}_1,\mathbb{X}_2)$ we can rewrite the the above as \cite{drummond_generalised_1980}
\begin{align}
    \partial_t P = \partial_i \left[ - \mathbb{A}_{i} + \partial_j \mathbb{D}_{ij} \right] P.
\end{align}
which allows us to identify $\mathbb{A}$ in \cref{app:eq:fw} as the drift field and similarly $\mathbb{D}$ as the diffusion tensor, both of which are analytic in the two components of $\mathbb{X}$. 

To arrive at the condition for detailed balance, the Fokker-Planck equation can be rewritten as a continuity equation $\partial_t P = \partial_i J_{i}$ with 
\begin{align}
    \label{ap:eq:J}
    J_{i}= -\mathbb{A}_{i}P + \partial_{j}(\mathbb{D}_{ij} P)
\end{align}
the probability current. The potential condition, equivalent to classical detailed balance \cite{gardiner_handbook_2009}, arises upon imposing that the probability current vanish $J_i = 0$. With this, upon rearranging terms in \cref{ap:eq:J} and assuming as before that $\mathbb{D}$ is invertible, we have (see also \cite{BartoloPRA_2016})
\begin{align}
    -\frac{\partial_i P}{P} = - \partial_i \log P = -\mathbb{D}_{i j}^{-1} \mathbb{A}_{j} + \mathbb{D}_{ij}^{-1} \partial_{k}\mathbb{D}_{jk}.
\end{align}
The condition above is equivalent to having the complex vector field in the last member of the equation be curl free, that is
\begin{align}
    \partial_j Z_i = \partial_i Z_j,\; Z_i \equiv  -\mathbb{D}_{i j}^{-1} \mathbb{A}_{j} + \mathbb{D}_{ij}^{-1} \partial_{k}\mathbb{D}_{jk}. \label{ap:eq:DB}
\end{align} 
Consequently putting $\partial_i U(\mathbb{X}) \equiv Z_i$, the steady-state complex-P function is $P \propto \exp(-U(\mathbb{X}))$.

In the case of interest in the main text, to obey \cref{ap:eq:smallDB}, we have to ensure that $\partial_{b_{\rm cl}} \frac{\overline{A}}{\overline{D}} = \partial_{\overline{b}_{\rm cl}} \frac{A}{D}$. Both of these quantities are equal to $2$, hence the time-reversed parametrization is a valid solution to the equations of motion. Moreover, the standard potential condition \cref{ap:eq:DB} obtained from the Fokker-Planck equation for the complex-P function is also satisfied. In the next section we detail the calculation of the extremized action using the time-reversed Ansatz and show that we recover the potential $U$ of the steady-state solution of the complex-P function up to sub-leading contributions.

For the case of dephasing, we would add a dissipator $ \mathbf{D}_{\ha^\dagger \ha}$. The corresponding drift and diffusion tensors are $\mathbb{A}_\phi = \begin{bmatrix}
    \mathbb{X}_1(1/2 -\mathbb{X}_1 \mathbb{X}_2)\\ \mathbb{X}_2 (1/2-\mathbb{X}_1\mathbb{X}_2)
\end{bmatrix}$ and $\mathbb{D}_\phi =\begin{bmatrix}
    0&1/2\\1/2&0
\end{bmatrix}$. We have that both potential conditions \cref{ap:eq:DB} and \cref{ap:eq:smallDB} are satisfied since $\partial_1 (\mathbb{D}^{-1}_\phi\mathbb{A}_\phi)_2 = \partial_2 (\mathbb{D}^{-1}_\phi\mathbb{A}_\phi)_1 = 1/4 - \alpha\beta$ and obviously $\partial \mathbb{D} = 0$. However adding two-photon dissipation and two-photon driving (see below \cref{ap:eq:deph2pho}) breaks both potential conditions. Note that the both conditions defined above require that $\mathbb{D}$ be invertible or at least the quantities $Z$ and $Z'$ to be well-defined, \ie that $\mathbb{A}$ is orthogonal to the kernel of $\mathbb{D}$. This is no longer the case for dephasing. Even in domains where $\mathbb{D}$ is invertible and we could locally define the vector fields entering the detailed-balance conditions \cref{ap:eq:smallDB,ap:eq:DB}, those conditions are broken.

\begin{table}
    \centering
    \begin{tabular}{|c|c|c|}
        \hline
         Liouvillian & \cref{ap:eq:DB} & \cref{ap:eq:smallDB}   \\
         \hline
         HTRS \cref{app:eq:Lindbladian_density} & Yes  & Yes\\
         \hline
         $\kappa_\phi\mathbf{D}_{\ha^\dagger\ha}+ \kappa_2 \mathbf{D}_{\ha^2-\alpha_0^2}$ & No & No\\
         \hline
         $\kappa_\uparrow\mathbf{D}_{\ha^\dagger} + \kappa_2 \mathbf{D}_{\ha^2-\alpha_0^2}$ & No & No\\
         \hline
    \end{tabular}
    \caption{Summary of the applicability of the conditions defined in \cref{app:sec:Time-R} for several systems.}
    \label{tab:my_label}
\end{table}

\section{Calculation of the action} \label{app:sec:Action}
In this appendix we detail the calculation of the extremized action and the prediction of the switching rate in the main text~[MT]. We start with the action in the rotated variables \cref{app:eq:canonical_transformation} and inject the time-reversed parametrization $\begin{bmatrix}
    \overline{b}_{\rm q}^c\\- b_{\rm q}^c
\end{bmatrix} = -\mathbb{D}^{-1}\mathbb{A}$,
\begin{align}
\begin{split}
    iS_{i\to j} &= \int \left(-\bar{b}_q^c\partial_t b_{cl} + b_q^c\partial_t \bar{b}_{cl}\right) \text{d}t,\\
    &=\int (\mathbb{D}^{-1}\mathbb{A}) \cdot \left(\partial_t\begin{bmatrix}
        b_{\rm cl}\\ \overline{b}_{\rm cl}
    \end{bmatrix}\right)\text{d} t
\end{split}
\end{align}
where $i,j$ are the distinct indices of the two fixed points $\alpha_1, \alpha_2$.
The boundary condition are such that when $t\to -\infty$, $b_{\rm cl}= \alpha_{i}$ and $b_{\rm q} = 0$ and when $t\to \infty$, $b_{\rm cl}= \alpha_{j}$ and $b_{\rm q} = 0$. As explained in the previous section, the validity of the time-reversed Ansatz derived in the previous section is equivalent to the path-independence of the above integral. We now consider the drift and diffusion tensors \cref{app:eq:HTRS_AD}. The boundary conditions are such that $\overline{b}_{\rm cl} = b_{\rm cl}^*$ and since the trajectory is path independent we choose $\overline{b}_{\rm cl} = b_{\rm cl}^*$ on the whole integration to obtain
\begin{align}
    iS_{i\to j} & = \int [2b_{\rm cl}^* + g(b_{\rm cl})] \text{d}b_{\rm cl}  + [2b_{\rm cl} + g(b_{\rm cl})^*]\text{d}b_{\rm cl}^*.
\end{align}
Where we have defined,
\begin{align}
\begin{split}
    {g(b_{\rm cl})} &= - \frac{2K_1b_{\rm cl} + 2{\Lambda_{3}}^* b_{\rm cl}^{2} + 2\Lambda_{1}}{K_2 b_{\rm cl}^{2} - 2 {\Lambda_{2}} - 2 \Lambda_{3} b_{\rm cl}}\\
    g(z)&= \frac{-2 \Lambda_3^*}{K_2} + \frac{Az +B}{K_2 (z-z_+)(z-z_-)}
\end{split}
\end{align}
with $A = -2K_1 - \frac{4|\Lambda_3|^2}{K_2}$ and $B=-2\Lambda_1 - \frac{4\Lambda_3^*\Lambda_2}{K_2}$ and $z_{\pm} = \frac{\Lambda_3}{K_2} \pm \frac{\sqrt{\Lambda_3^2 + 2 \Lambda_2K_2}}{K_2}$. 

This integral that can be separated in two parts. The first part gives $\left[ 2|b_{cl}|^2\right]^{b_{\rm cl} = \alpha_u}_{b_{\rm cl} = \alpha_{i}}$. To integrate the second part of the integral, we decompose $g$ in partial fractions and obtain,
\begin{align}
    g(z) = \frac{-2 \Lambda_3^*}{K_2} + \frac{C_+}{K_2 (z-z_+)} + \frac{C_-}{K_2 (z-z_-)}
\end{align}
where $C_\pm = \frac{Az_\pm + B}{z_\pm - z_\mp}$. We can now look for a potential function, a function such that $g(z) \text{d}z + {g(z)}^* \text{d}{z}^* = \text{d}\Psi$. The following function satisfies this condition up to a constant,
\begin{align}
\begin{split}
    \Psi(z) =& 2\Re\left(\frac{-2 \Lambda_3^* z}{K_2} \right)\\
    & + 2\Re\left[\frac{C_+}{K_2} \ln\left(z-z_+ \right) \right]\\
    & + 2\Re\left[\frac{C_-}{K_2} \ln\left(z-z_- \right)\right],
\end{split}
\end{align}
where $\ln(z)$ is the complex logarithm. Note that the complex logarithm is multi-valued, this has no impact on the value of the integral $\int \text{d}\Psi$. Note that the choice of the branch cut of the complex logarithm impacts the prediction of the switching rate. We chose the latter to be on the semi-infinite line $i]-\infty,0]$, so to agree with numerical diagonalization in the case $\Lambda_3\neq0$.

Finally we can write the action in a closed form,
\begin{align}\label{app:eq:Swicthing_rate}
    iS_{i\to j} = \left[ \Phi(b_{\rm cl})\right]^{b_{\rm cl} = \alpha_u}_{b_{\rm cl} = \alpha_i},
\end{align}
with $\Phi(z) = 2|z|^2 + \Psi(z)$.
The rates are then given by $\Gamma_{i\to j} \propto e^{iS_{i\to j}}$. Note that in the limit $\Lambda_2 \gg \Lambda_1, \Lambda_2, \Delta, \kappa_1,\kappa_2$ we have that $\Phi(\alpha_{ss})$ scales as $\sqrt{\Lambda_2}$ and is hence sub-leading compared to $|\alpha_{ss}|^2$.

We have effectively calculated the potential of $Z' = \mathbb{D}^{-1}\mathbb{A}$. We can also calculate the potential of the complex-P function defined with $Z_i = \mathbb{D}^{-1}_{i,j} (\mathbb{A}_j - \partial_k \mathbb{D}_{k,j})$. The additional term $ \frac{\partial_\alpha D}{D} + \frac{\partial_\beta \overline{D}}{\overline{D}}$ has the potential function $\ln( D \overline{D})$ and results in a prefactor to the potential obtained above. We have,
\begin{align}
    P\propto \frac{\exp[\bar{\Phi}(\alpha,\beta)]}{D\overline{D}},
\end{align}
where $\Phi(z) = \bar{\Phi}(z,z^*)$. The normalization has to be computed on a bounded contour encircling the singularities for both variables $\alpha$ and $\beta$ (see~\cite{drummond_generalised_1980, BartoloPRA_2016} for a detailed discussion). We have checked that the system studied in the main text satisfies a detailed balance condition for the complex-P function, equivalently obeying HTRS. Moreover we have showed how the potential of the steady-state complex-P function is related to the exponent entering the switching rate.

\section{Numerical integration of the equations of motion for dephasing}
\label{app:sec:numint}

For dephasing, two-photon loss, and two-photon drive, the Liouvillian reads,
\begin{align} \label{ap:eq:deph2pho}
    \mathbf{L} = \kappa_\phi\mathbf{D}_{\ha^\dagger\ha}+ \kappa_2 \mathbf{D}_{\ha^2-\alpha_0^2}
\end{align}
The saddle-point equations of motion for the variables $\left[b_{\rm cl}, \overline{b}_{\rm cl}, b_{\rm q},\overline{b}_{\rm q}\right] $ read
\begin{align}
    \begin{split}
        \partial_t{b}_{\rm cl} = & \kappa_{2} \left[(\bar{b}_{\rm q}+\bar{b}_{\rm cl}) (\alpha_0^{2} -b_{\rm cl}^{2})\right]\\ &+ \kappa_{\phi} b_{\rm cl}\left[-\bar{b}_{\rm q}b_{\rm cl} - b_{\rm q} \bar{b}_{\rm cl} - 1/2\right], \\
        \partial_t{\bar{b}}_{\rm cl} = & \kappa_{2} \left[(b_{\rm cl}-b_{\rm q})(\alpha_0^2-\bar{b}_{\rm cl}^2)\right]\\ & + \kappa_{\phi} \bar{b}_{\rm cl}\left[\bar{b}_q b_{cl}  + \bar{b}_{cl} b_{q} - 1/2\right], \\
        \partial_t{{b}}_{\rm q} = & \kappa_{2} \left[\bar{b}_q( \alpha_0^{2} - b_{cl}^{2}) + 2 \bar{b}_{cl} b_{cl} b_{q} - \bar{b}_{cl} b_{q}^{2} \right] \\  &+ \kappa_{\phi} \left(- \bar{b}_q b_{cl} b_{q} - \bar{b}_{cl} b_{q}^{2} + b_{q}/2 - b_{\rm cl}\right),\\
        \partial_t{\bar{b}}_{\rm q} = &\kappa_{2} \left[- \bar{b}_q^{2} b_{cl} - 2 \bar{b}_q \bar{b}_{cl} b_{cl} + (\bar{b}_{cl}^{2} - \alpha_0^{2})b_{q}\right]\\ & + \kappa_{\phi} \left(- \bar{b}_q^{2} b_{cl} - \bar{b}_q \bar{b}_{cl} b_{q} - \bar{b}_q/2-\bar{b}_{\rm cl}\right).
    \end{split}
\end{align}

As noted in previous works~\cite{kamenev_field_2011, thompson_qubit_2022}, $b_q^*$ (the complex conjugate of $b_q$) and $\bar{b}_q$ do not satisfy the same equations of motion. This means that to reach the saddle points in the path integral, the integration contour for $b_q$ must be deformed. We chose the contour $\bar{b}_{\rm q} = -b_{\rm q}^*$ and fix this feature by making the replacement $\tilde b_q = i b_q$~\cite{kamenev_field_2011, lee_arealtime_2024}. 
The equations of motion satisfied by the new variables are complex conjugate. 

In the noise-free subspace where $b_{\rm q} =0$, we have three fixed points, $b_{\rm cl}=0$ and $b_{\rm cl} = \pm\sqrt{\alpha_0^2 - \frac{\kappa_\phi}{2\kappa_2}}$. Since the dynamics are generated by a time-independent ``energy" $\mathcal{L}$, every fixed point is a saddle point and lies at the intersection of two planes: an ``attractive" 2D plane tangent to trajectories converging towards the fixed point, the noise-free plane, and another ``repulsive" 2D plane tangent to trajectories leaving the fixed points. To find these two planes, we compute the Jacobian at the fixed stable points, and find its eigenvalues or Lyapunov exponents. The repulsive plane is spanned by the eigenvectors associated to eigenvalues with positive real part. 

To find the instanton paths, we initialize the system on the repulsive plane while staying as close as possible to the fixed point. We sweep the initial vector on this plane and select the two trajectories that pass the closest through the fixed point $\left[ 0,0,0,0\right]$. 

\section{Extension to many-body systems}
In this appendix, we demonstrate the generality of our method by extending the calculation of switching rates to a many-body system. Furthermore, we provide a step-by-step methodology for estimating these rates using the time-reversal Ansatz, serving as a practical guide for applying the result presented in the main text. Specifically, this appendix focuses on a sub-model of the many-body system satisfying HTRS introduced in \citealt{roberts_competition_2023}.

\textit{Step $1$: From the Lindbladian to the Keldysh action}. We begin by defining the dynamics of the system using the Lindblad master equation. The dynamics are governed by the Lindbladian $\mathbf{L}(\hat{\rho})$
\begin{align}\label{app:eq:System}
\begin{split}
    \mathbf{L} (\hat{\rho})&= -i \left[\hat{H},\hat{\rho}\right] + \kappa \sum_i \mathbf{D}_{\hat{a}_i}(\hat{\rho})\\
    \hat{H} &= \Delta\sum_i \hat{n}_i -\frac{K}{2}\left(\sum_i \hat{n}_i\right)^2 + \sum_{i>j} \left(M_{i,j} \ha^\dagger_i \ha^\dagger_j +\text{h.c.} \right).
\end{split}
\end{align}
Here $\hat{n}_i = \hat{a}^\dagger_i \hat{a}_i$ is the number operator of the mode $i$. Before deriving the Keldysh action we note that this system can be simplified with a Bogolioubov-de Gennes transformation. 

We introduce the unitary matrix $S$ and define the new bosonic operators
\begin{align}
    \begin{split}
    \hat{b}_i = S_{i,j}\hat{a}_j,
    \end{split}
    \begin{split}
        \hat{b}_i^\dagger = S_{i,j}^* \hat{a}_j^\dagger.
    \end{split}
\end{align}
The unitarity of $S$ ensures both the canonical commutation relations, $\left[\hat{b}_i, \hat{b}^\dagger_j\right] = \delta_{i,j}$, and the conservation of the number and dissipation terms: $\sum_i \hat{a}_i^\dagger \ha_i = \sum_i \hb_i^\dagger \hb_i$ and $\sum_i \mathbf{D}_{\ha_i} = \sum_i \mathbf{D}_{\hb_i}$. This yields the transformed Lindbladian
\begin{align}
    \mathbf{L} (\hat{\rho})&= -i \left[\hat{H},\hat{\rho}\right] + \kappa_1 \sum_i \mathbf{D}_{\hat{b}_i}(\hat{\rho})\\
    \hat{H} &= \Delta\sum_i \hat{n}_i -\frac{K}{2}\left(\sum_i \hat{n}_i\right)^2 + \sum_{i>j} \left(N_{i,j} \hb^\dagger_i \hb^\dagger_j +\text{h.c.} \right),
\end{align}
where the transformed coupling matrix is $N_{i,j} = S_{l,i}^* M_{l,m} S_{m,j}^*$, or $N= (S^*)^TM (S^*)$ in matrix notation. Since $M$ is a complex symmetric matrix, the Autonne–Takagi factorization guarantees the existence of a unitary matrix, $U^\dagger U= \mathbb{I}$, such that $U^T M U$ is a real diagonal matrix with non-negative entries. This confirms that, without loss of generality, we can choose the original matrix $M$ in \cref{app:eq:System} to be a real diagonal matrix, $M_{i,j} = \delta_{i,j} G_i$.

Applying the Keldysh path integral formalism to the simplified Lindbladian $\mathbf{L}(\hat{\rho})$ then yields the Lindbladian density
\begin{align}
\begin{split}
    \mathcal{L}[\mathbf{a}_l, \mathbf{a}_r] &= -i\tilde{\Delta} (\bar{\mathbf{a}}_l^T \mathbf{a}_l - \mathbf{a}_r^T \bar{\mathbf{a}}_r) + \frac{iK}{2}\left[(\bar{\mathbf{a}}_l^T \mathbf{a}_l)^2 - (\mathbf{a}_r^T \bar{\mathbf{a}}_r)^2\right] \\
    &- i \left[\bar{\mathbf{a}}_l^{T}\mathbf{G} \bar{\mathbf{a}}_l + \mathbf{a}_l^{T}\mathbf{G} {\mathbf{a}}_l- \bar{\mathbf{a}}_r^{T}\mathbf{G} \bar{\mathbf{a}}_r - \mathbf{a}_r^{T}\mathbf{G} \mathbf{a}_r  \right]\\
    &+ \kappa \left[\mathbf{a}_l^{T} \bar{\mathbf{a}}_r - \frac{1}{2} \left(\bar{\mathbf{a}}_l^T \mathbf{a}_l + \mathbf{a}_r^T \bar{\mathbf{a}}_r\right)\right],
    \end{split}
\end{align}
where $\bar{\mathbf{a}}_l^T = \begin{bmatrix}
    \bar{a}_l^{(1)}, \bar{a}_l^{(2)}, \dots, \bar{a}_l^{(n)}
\end{bmatrix}$ and $\mathbf{a}_r^T = \begin{bmatrix}
    a_r^{(1)}, a_r^{(2)},\dots, a_r^{(n)}
\end{bmatrix}$, $\tilde{\Delta} = \Delta - K/2$, and $a_l^{(i)}$ and $a_r^{(i)}$ are the fields associated with the annihilation operator $\hat{a}_i$ acting on the left and right of the density matrix, respectively.

We now perform the Keldysh rotation together with the change of variable defined in \cref{app:eq:canonical_transformation} on each mode. This transformation moves from the left $a_l^{(i)}$ and right $a_r^{(i)}$ fields to the classical $a_{\rm cl}^{(i)}$ and quantum $a_{\rm q}^{(i)}$ fields
\begin{align}
    \begin{split}
        \mathbf{b}_{\rm cl} = \mathbf{a}_l,
    \end{split}
    ~
    \begin{split}
        \bar{\mathbf{b}}_{\rm cl} = \bar{\mathbf{a}}_r,
    \end{split}
    ~~
    \begin{split}
        \mathbf{b}_{\rm q} = \mathbf{a}_l - \mathbf{a}_r,
    \end{split}
    ~
    \begin{split}
        \bar{\mathbf{b}}_{\rm q} = \bar{\mathbf{a}}_l - \bar{\mathbf{a}}_r.
    \end{split}
\end{align}

Substituting these definitions into the Lindbladian density $\mathcal{L}[\mathbf{a}_l, \mathbf{a}_r]$ and expanding the terms yields the Lindbladian density in the classical-quantum basis
\begin{align}
    \begin{split}
    \mathcal{L}[ \mathbf{b}_{\rm cl }, \mathbf{b}_{\rm q} ] &= (-i\tilde{\Delta})  \left[\bar{\mathbf{b}}_{\rm q}^T \mathbf{b}_{\rm cl} 
     + \mathbf{b}_{\rm q} ^T \bar{\mathbf{b}}_{\rm cl}\right] \\
    &+ \frac{iK}{2}\left[\bar{\mathbf{b}}_{\rm q}^{2T}\mathbf{b}_{\rm cl}^2 + 2 \bar{\mathbf{b}}_{\rm q}^T\bar{\mathbf{b}}_{\rm cl}^T \mathbf{b}_{\rm cl}^2
    -\mathbf{b}_{\rm q }^{2T}\bar{\mathbf{b}}_{\rm cl}^2 + 2\mathbf{b}_{\rm q}^T \mathbf{b}_{\rm cl }^T \bar{\mathbf{b}}_{\rm cl}^2\right] \\
    &-i\left[\bar{\mathbf{b}}_{\rm q}^T \mathbf{G}\bar{\mathbf{b}}_{\rm q} + 2\bar{\mathbf{b}}_{\rm q}^T \mathbf{G}\bar{\mathbf{b}}_{\rm cl} + 2 \mathbf{b}_{\rm q}^T \mathbf{G} \mathbf{b}_{\rm cl} - \mathbf{b}_{\rm q}^T\mathbf{G} \mathbf{b}_{\rm q} \right]\\
    &- \frac{\kappa}{2} \left[\bar{\mathbf{b}}_{\rm q}^T\mathbf{b}_{\rm cl} - \mathbf{b}_{\rm q}^T \bar{\mathbf{b}}_{\rm cl}\right],
    \end{split}
\end{align}

As in \cref{app:sec:Time-R}, the Lindbladian density can be rewritten as a quadratic Hamiltonian, with position $\mathbb{X}^T = \begin{bmatrix}
    \mathbf{b}_{\rm cl}^T, \bar{\mathbf{b}}_{\rm cl}^T
\end{bmatrix}$ and the conjugate momentum $\mathbb{P}^T = \begin{bmatrix}\bar{\mathbf{b}}_{\rm q}^T, -\mathbf{b}_{\rm q}^T
\end{bmatrix} $. The Lindbladian density takes the form
\begin{align}
    \begin{split}
    \mathcal{L}[ \mathbf{b}_{\rm cl }, \mathbf{b}_{\rm q} ] &= \mathbb{P}^T \mathbb{A}_N + \mathbb{P}^T \mathbb{D}_N \mathbb{P}
    \end{split}
\end{align}
The drift vector $\mathbb{A}_N$ and the diffusion tensor $\mathbb{D}_N$ are identified as
\begin{align}
\begin{split}
    \mathbb{A}_N &= \begin{bmatrix}
        (-i\tilde{\Delta}-\frac{\kappa}{2})\mathbf{b}_{\rm cl} +iK\bar{\mathbf{b}}_{\rm cl}^T \mathbf{b}_{\rm cl}^2 
        - i 2\mathbf{G}\bar{\mathbf{b}}_{\rm cl}\\
        (i\tilde{\Delta}-\frac{\kappa}{2})\bar{\mathbf{b}}_{\rm cl} - iK{\mathbf{b}}_{\rm cl}^T \bar{\mathbf{b}}_{\rm cl}^2 + i 2\mathbf{G}{\mathbf{b}}_{\rm cl}
    \end{bmatrix}
    \\
    \mathbb{D}_N &= \begin{bmatrix}
        -i\mathbf{G}+ i\frac{K}{2} \mathbf{b}_{\rm cl}{\mathbf{b}}_{\rm cl}^T & 0\\
        0& i\mathbf{G} - i\frac{K}{2} \bar{\mathbf{b}}_{\rm cl}\bar{\mathbf{b}}_{\rm cl}^T 
    \end{bmatrix}
    \end{split}
\end{align}

We have successfully obtained the Keldysh action and can now define the drift vector and diffusion tensor. The next step is to check the validity of the time-reversed Ansatz \cref{ap:eq:smallDB}.

\textit{Step 2: Checking the potential condition}. The central requirement for using the time-reversed Ansatz is that the potential condition is satisfied. This condition is equivalent to requiring the vector field $ \mathbb{Z}_N = \mathbb{D}_{N}^{-1} \mathbb{A}_N$ to be such that $\partial_j \mathbb{Z}_i = \partial_i \mathbb{Z}_j$. The satisfaction of this condition is what guarantees the validity of the time-reversed Ansatz $\mathbb{P}^c = - \mathbb{D}_{N}^{-1} \mathbb{A}_N$. The potential condition on $\mathbb{Z}_N$ ensures that the parametrization $\mathbb{P}=\mathbb{P}^c$ defines a solution to the equations of motion of both the fields .$\mathbf{b}_{\rm cl}$ and $\mathbf{b}_{\rm q}$ in which $\mathbf{b}_{\rm cl}$ follows the time-reversed counterpart of the equations of motion where $\mathbf{b}_{\rm q} = 0$.

We assume the diagonal matrix G (containing the two-photon drive strengths) is invertible. (If $\mathbf{G}_i=0$, the dynamics must be analyzed in a lower-dimensional submanifold where $\mathbf{b}_{\rm q}^{(i)}=0$, similar to handling singular diffusion tensors in Fokker-Planck equations \cite{gardiner_handbook_2009}).

The matrix $\mathbb{D}_N$ is invertible if and only if, $1 - \frac{K}{2}\mathbf{b}_{\rm cl}^T \mathbf{G}^{-1}\mathbf{b}_{\rm cl} \neq 0$ and $1 - \frac{K}{2}\bar{\mathbf{b}}_{\rm cl}^T \mathbf{G}^{-1}\bar{\mathbf{b}}_{\rm cl} \neq 0$. Using the Sherman-Morisson formula we obtain the inverse of $\mathbb{D}_N$,
\begin{align}
    \mathbb{D}_N^{-1} = \begin{bmatrix}
        i \mathbf{G}^{-1} + i \frac{\mathbf{G}^{-1}\mathbf{b}_{\rm cl}\mathbf{b}_{\rm cl}^T \mathbf{G}^{-1}}{\frac{2}{K} -\mathbf{b}_{\rm cl}^T \mathbf{G}^{-1}\mathbf{b}_{\rm cl}} &0\\
        0& -i \mathbf{G}^{-1} - i \frac{\mathbf{G}^{-1}\bar{\mathbf{b}}_{\rm cl}\bar{\mathbf{b}}_{\rm cl}^T \mathbf{G}^{-1}}{\frac{2}{K} -\bar{\mathbf{b}}_{\rm cl}^T \mathbf{G}^{-1}\bar{\mathbf{b}}_{\rm cl}}
    \end{bmatrix} .
\end{align}
We then substitute this into the definition of the vector field $\mathbb{Z}_N$
\begin{align}
    \mathbb{Z}_N &= \begin{bmatrix}
        2\bar{\mathbf{b}}_{\rm cl} +  (\tilde{\Delta} -i\kappa/2) \frac{\mathbf{G}^{-1}\mathbf{b}_{\rm cl}}{1-\frac{K}{2} \mathbf{b}_{\rm cl}^T \mathbf{G}^{-1} \mathbf{b}_{\rm cl}}\\
        2{\mathbf{b}}_{\rm cl} +  (\tilde{\Delta} +i\kappa/2) \frac{\mathbf{G}^{-1}\bar{\mathbf{b}}_{\rm cl}}{1-\frac{K}{2} \bar{\mathbf{b}}_{\rm cl}^T \mathbf{G}^{-1} \bar{\mathbf{b}}_{\rm cl}}
    \end{bmatrix}.
\end{align}
By inspection of the components of $\mathbb{Z}_N$, we confirm that they satisfy the required cross-derivative conditions, $\partial_i (\mathbb{Z}_N)_j = \partial_j (\mathbb{Z}_N)_i$. The potential condition is satisfied, thereby validating the use of the time-reversed Ansatz for this multi-mode system.

\textit{Step 3: Time-reversal Ansatz and analytical extremization of the action}
The satisfaction of the potential condition in Step 2 validates the use of the time-reversed Ansatz to find the instanton path that extremizes the Keldysh action.

We choose the specific path parametrization for the quantum fields $\mathbb{P} = \mathbb{P}^c$. As we noted in Step 2, this choice ensures that the resulting equations of motion lead to a path where the classical field $\mathbf{b}_{\rm cl}$ follows the time-reversed counterpart of the noise-free equations of motion.

By inserting the Ansatz $\mathbb{P} = \mathbb{P}^c$ back into the Keldysh action,
the full action simplifies into a total derivative, enabling explicit integration. The exponential factor of the switching rate is then determined by the potential function $\Phi_N$
\begin{align}
\begin{split}
    iS &= \int (\mathbb{D}_N^{-1} \mathbb{A}_{N})^T \begin{bmatrix}
        \partial_t \mathbf{b_{cl}}\\
        \partial_t \bar{\mathbf{b}}_{cl}
    \end{bmatrix} \text{d}t,\\
    iS&=\int \text{d} \Phi_N,
\end{split}
\end{align}
The potential function $\Phi_N(\mathbf{b}_{\rm cl}, \bar{\mathbf{b}}_{\rm cl})$ for the multi-mode system is explicitly calculated to be
\begin{align}
\begin{split}
    \Phi_N(\mathbf{b}_{\rm cl}, \bar{\mathbf{b}}_{\rm cl}) &=  2 \bar{\mathbf{b}}_{\rm cl}^T \mathbf{b}_{\rm cl} \\
    &+ 2 \Re\left[\frac{-\tilde{\Delta}+i\kappa/2}{K} \ln \left(\mathbf{b}_{\rm cl}^T \mathbf{G}^{-1}\mathbf{b_{\rm cl}}- 2 /K \right) \right].
\end{split}
\end{align}

\textit{Step 4: Estimating the Switching Rate and Dissipative Gap}
The final step is to use the potential function $\Phi_N(\mathbf{b}_{\rm cl}, \bar{\mathbf{b}}_{\rm cl})$ to determine the exponential factor of the switching rate.

The first requirement is that the system must exhibit metastability. This occurs when the noise-free (classical) equations of motion, $\partial_t \mathbb{X} = \mathbb{A}_N(\mathbb{X})$, admit multiple stable fixed points, $\mathbb{X}_{ss}$. In this context, the fixed points satisfy $\mathbb{A}_N(\mathbb{X}) =0$

For the bistable systems studied in the main text we have three fixed points. One fixed point $\mathbf{0} = \begin{bmatrix}
    0\\0
\end{bmatrix}$ is unstable in the noise-free subspace, two others are stable $\mathbb{X}_1,~\mathbb{X}_2$. As in the previous calculation \cref{app:eq:Swicthing_rate}, the switching path follows the time-reversed parametrization from $\mathbb{X}_1$ to $\mathbf{0}$, then the path follows the noise-free subspace from $\mathbf{0}$ to $\mathbb{X}_2$. The second part of the path accumulates no action and the switching rate can be expressed as
\begin{align}
    \Gamma_{1\to 2} &= \exp\left[\Phi_N(\mathbf{0}) - \Phi_n(\mathbb{X}_1)\right].
\end{align}

For a bistable system, the dissipative gap ($\Gamma$) is directly given by the average of the forward and reverse switching rates
\begin{align}
    \Gamma = \frac{\Gamma_{1\to 2} + \Gamma_{2\to 1}}{2}.
\end{align}

The methodology presented here could be extended to multi-stable systems. The primary difference is that the individual switching rates $\Gamma_{i\to j}$ must be calculated for every transition path. These rates then form the entries of a transition rate matrix, the dissipative gap ($\Gamma$) for a multi-stable system is then determined by the magnitude of the smallest non-zero eigenvalue of this transition rate matrix~\cite{norris_markov_1997}.

\begin{figure}[t!]
    \centering
    \includegraphics[width=\linewidth]{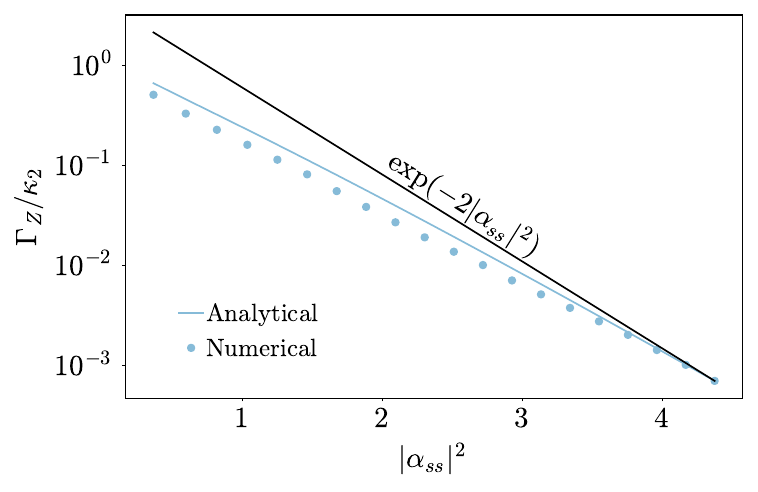}
    \caption{Dissipative gap of the two-mode Kerr-coupled and two-photon driven system. In this system one mode is frozen to vacuum while the second mode exhibits bistability. The dissipative gap is obtain with the switching probability from one fixed point to the other. The prefactor to the rate in \cref{app:eq:rate_BH} is adjusted so that the numerical and analytical prediction intersect on the largest $|\alpha_{ss}|^2$ point. The dimensionless parameters are $K=1,~\kappa =1,~ \Delta=0,~\mathbf{G}_1 = 0.2$ and $\mathbf{G}_0$ is swept from $0.5$ to $2.45$.}
    \label{fig:Dimer_biftlip}
\end{figure}

\textit{Parameter regime for bistability} We now detail the parameter regime in which the system described by \cref{app:eq:System} exhibits bistability at the level of the equations of motion. This allows for an explicit estimation of the dissipative gap by calculating the switching probability, as defined in \cref{app:eq:Swicthing_rate}.

The noise-free equations of motion are obtained by setting the quantum fields to zero ($\mathbf{b}_{\rm q} =0$)
\begin{align}
    \begin{bmatrix}
        \partial_t \mathbf{b}_{\rm cl}\\
        \partial_t \bar{\mathbf{b}}_{\rm cl}
    \end{bmatrix} &= \mathbb{A}_N.
\end{align}
Therefore, fixed points $\mathbf{b}_{\rm cl} = \mathbf{z}$ satisfy the condition $\mathbb{A}_N=0$,
\begin{align}
    \left(\tilde{\Delta} -i\frac{\kappa}{2} -K \bar{\mathbf{z}}^T\mathbf{z} \right) \mathbf{z} + 2 \mathbf{G}\bar{\mathbf{z}} =0.
\end{align}
One solution is $\mathbf{z} = 0$. Introducing the polar notation $\mathbf{z}_i = r_i e^{i\theta_i}$, the non-zero solutions satisfy the two coupled conditions
\begin{align}
\begin{split}
    \left(\tilde{\Delta} -K \sum_ir_i^2 \right)  + 2 \mathbf{G}_i  \cos(2\theta_i) &=0,\\
     \sin(2\theta_i) &= -\frac{\kappa}{4 \mathbf{G}_i}.
\end{split}
\end{align}
The second equation implies that $\mathbf{b}_i$ has a non-zero solution only if the two-photon drive exceeds the decay rate $\mathbf{G}_i>\kappa/4$. If we assume non-degenerate coupling ($\mathbf{G}_i\neq\mathbf{G}_j$ for $i\neq j$), the first equation implies that for non-zero fixed points only a single mode is occupied ($\mathbf{b}_j =0$ for $j\neq i$). Finally the possible fixed points are given by $r_i^2 = \tilde{\Delta}/K + \sqrt{4\mathbf{G}_i^2/K^2 - \kappa^2/4K^2}$ which give rise to $\mathbf{b}_i = \pm \alpha_i$ (cat-like states).

To obtain a simple bistable regime, we specify the system size to $N=2$ (a dimer) and choose parameters such that only one mode is active: we set $\mathbf{G}_0>\kappa/4$ (active mode) and $\mathbf{G}_1<\kappa/4$ (frozen mode). This regime yields three fixed points: the vacuum $\mathbf{0}=(0,0)$ and the two cat-states $\mathbb{X}_0 = (\alpha_0,0)$ and $-\mathbb{X}_0=(-\alpha_0,0)$, where $\alpha_0$ is the steady-state amplitude of the active mode.

We express the dissipative gap ($\Gamma_Z$) by using the switching probability between both fixed points $\pm \mathbb{X}_0$. Since the rates are symmetric, $\Gamma_{\mathbb{X}_0\to -\mathbb{X}_0} = \Gamma_{-\mathbb{X}_0\to \mathbb{X}_0}$, we obtain that the dissipative is
\begin{align}\label{app:eq:rate_BH}
    \Gamma_{Z} \propto \exp\left[ \Phi_2(\mathbf{0}) - \Phi_2(\mathbb{X}_0) \right].
\end{align}

In \cref{fig:Dimer_biftlip}, we compare the analytical prediction of \cref{app:eq:rate_BH} with the numerical diagonalization of the Lindbladian for this two-mode system. The dimensionless parameters are fixed to $K=1,~\kappa =1,~ \Delta=0,~\mathbf{G}_1 = 0.2$, while the active mode coupling $\mathbb{G}_0$ is swept from $0.5$ to $2.45$ (which corresponds to sweeping the steady-state amplitude $|\alpha_{ss}|$). The numerical diagonalization confirms the predicted exponential scaling across the parameter range. Crucially, in the large amplitude limit, the numerical computation asymptotically approaches the dominant, generic scaling $\Gamma_Z \propto e^{-2|\alpha_{ss}|^2}$, which is consistent with the other examples presented in the main text.

\bibliographystyle{apsrev4-2}
\bibliography{ref}